\documentclass[a4paper,fleqn]{cas-sc}

\def\tsc#1{\csdef{#1}{\textsc{\lowercase{#1}}\xspace}}
\tsc{WGM}
\tsc{QE}
\usepackage{amsmath}
\usepackage{graphicx}
\usepackage{caption}
\usepackage{url}
\usepackage{subcaption}
\usepackage[numbers,sort&compress]{natbib}
\DeclareMathOperator{\sgn}{sgn}   
\DeclareMathOperator{\sech}{sech}
\DeclareMathOperator{\artanh}{artanh}

\newcommand{\Ups}{\Upsilon}
\newcommand{\Om}{\tilde\Omega}
\newcommand{\Del}{\tilde\Delta}

\begin{document}
\let\WriteBookmarks\relax
\def\floatpagepagefraction{1}
\def\textpagefraction{.001}

\shorttitle{}    

\shortauthors{}  

\title [mode = title]{Strongly chirped dissipative solitons in normal and anomalous dispersion regimes}  

\author[1]{V. L. Kalashnikov}[orcid=0000-0002-3435-2333]

\cormark[1]


\ead{vladimir.kalashnikov@ntnu.no}

\ead[url]{https://www.ntnu.edu/employees/vladimir.kalashnikov}

\credit{Conceptualization, Methodology, Software, Formal analysis, Visualization, Data Curation, Writing - Original Draft}

\affiliation[1]{organization={Department of Physics, Norwegian University of Science and Technology},
            addressline={Realfagbygget, Gl{\o}shaugen, H{\o}gskoleringen}, 
            city={Trondheim},
            postcode={NO-7491},
            country={Norway}}

\author[2,3]{E. Sorokin}


\ead{evgeni.sorokin@tuwien.ac.at}


\credit{Conceptualization, Writing - Review \& Editing}

\affiliation[2]{organization={Institut f{\"u}r Photonik, TU Wien},
            addressline={Gu{\ss}hausstra{\ss}e 27/387}, 
            city={Vienna},
            postcode={A-1040},
            country={Austria}}

\affiliation[3]{organization={ATLA lasers AS},
            addressline={Richard Birkelands vei 2B}, 
            city={Trondheim},
            postcode={NO-7034},
            country={Norway}}

\author[1]{A. Rudenkov}


\ead{alexander.rudenkov@ntnu.no}


\credit{Conceptualization}

\author[1,3]{I. T. Sorokina}


\ead{irina.sorokina@ntnu.no}


\credit{Supervision, Funding acquisition, Writing - Review \& Editing}

\cortext[1]{Corresponding author}



\begin{abstract}
We study strongly chirped dissipative solitons (DSs) of the cubic--quintic complex Ginzburg--Landau equation (CGLE) in both normal- and anomalous-group-delay dispersion regimes (NGD and AGD). Using a stationary-phase (adiabatic) approximation, we obtain analytic NGD and AGD spectra, allowing us to represent the DS parametric space as a master diagram that connects ratios of spectral filter parameters, group-delay dispersion (GDD), and self-phase (SPM) and self-amplitude (SAM) modulations to a scaled DS energy. A central result is that dissipative-soliton resonance (DSR) emerges in the NGD and follows directly from the admissibility constraints, whereas, in the AGD, it appears only when the DSR locus intersects the adiabatic existence window. The last requires sufficiently strong quintic (saturable) SPM.

In the NGD, the spectra are inherently truncated. In the AGD, the spectral envelopes are approximately self-similar in an energy-scaling regime, have two-horns, smooth wings, and frequency truncation is provided by a spectral dissipation. In this regime, the energy dependence is captured predominantly by a scalar prefactor, while the unit-peak spectral core remains nearly invariant.

We demonstrate that, in AGD, there is a separation between the absolute first-order autocorrelation magnitude and the normalized coherence shape, revealing two distinct correlation times: a short scale set by the effective spectral dissipation bandwidth and a long scale controlled by the intrinsic spectral core, as in NGD.

Finally, we outline a thermodynamic interpretation of the correlation-scale separation and its connection to an entropy-driven transition between single- and multi-soliton states, thereby constraining the DS energy scalability. We also discuss how the same two-scale coherence phenomenology can be relevant to weakly dissipative Bose-Einstein condensate (BEC) settings (atomic and driven optical condensates) in the presence of bandwidth-limited spectral and/or linear and nonlinear losses.
\end{abstract}


\maketitle

\section{Introduction}\label{intro}

Traditionally, the mathematically purest form of a soliton \cite{PhysRevLett.15.240}, i.e., a localized wave packet exhibiting perfect elastic scattering, is found in integrable systems like the nonlinear Schrödinger, KdV, or 1D non-trap Gross-Pitaevskii equations \cite{dodd1982solitons}. However, this term is now also widely used for stable solitary waves in non-integrable systems, where a balance between linear and nonlinear effects allows them to propagate robustly. It is very interesting, both experimentally and theoretically, that solitary waves can also exist in dissipative systems with gain and loss far from thermodynamic equilibrium. These localized structures, which rely on a continuous energy exchange with their environment for stability, were termed dissipative solitons (DSs) \cite{ankiewicz2008dissipative},
and found a practical application, in particular, for the generation of energy-scalable ultrashort pulses from mode-locked lasers \cite{grelu2012dissipative}. Mode-locked lasers have become an excellent testbed for DS theory, demonstrating phenomena such as soliton molecules \cite{PhysRevLett.79.4047,hause2008binding,soto2025vibrational}, pulsating/breathing solitons \cite{soto2000pulsating,lucas2017breathing,Peng31122025}, explosive instabilities \cite{soto2000pulsating,cundiff2002experimental,grelu2012dissipative}, “soliton rains” \cite{grelu2012dissipative,PhysRevA.81.063829}, and composite solitons \cite{soto2002composite,kochetov2019logic} that have no counterparts in conservative soliton physics.

Unlike a non-dissipative soliton, whose energy is fixed for given parameters, a DS can exist over a broad range of energies and durations by adjusting the balance of gain/nonlinear loss and dispersion/phase nonlinearity \cite{grelu2012dissipative}, forming a key parameter characterizing the DS parametric space \cite{kalashnikov2006chirped}. Early theoretical formalisms for describing DS were based on a master equation – essentially a CGLE that includes terms for gain saturation, SAM, GDD, and SPM \cite{haus1975theory}. Soliton-like solutions of this master equation correspond to stable mode-locked pulses in a laser. By the mid-1990s, the detailed analysis of the cubic-quintic CGLE was performed, revealing that it admits stable pulsed solutions in both the AGD and NGD regimes \cite{kivshar1989dynamics,haus1991structures,akhmediev1995novel,akhmediev1996singularities,soto1997pulse}, spatial and spatio-temporal solitons \cite{malomed2005spatiotemporal,skarka2006stability}, solitons in BEC \cite{sakaguchi2003stable}, metamaterials and plasmonics \cite{smirnova2015dissipative}, and PT-symmetric systems \cite{he2013lattice}. In parallel, closely related coarse-grained descriptions arise for driven open condensates, where the long-wavelength phase dynamics can be mapped
to effective stochastic field theories and used to predict universal scaling of coherence and correlations in space and time \cite{sieberer2025universality}.

The interesting kind of stable DS is one that exhibits an instantaneous phase change within its slowly varying envelope. This phenomenon (chirp) usually develops in the NGD \cite{proctor1993characterization}. The presence of the chirp causes the energy fluxes inside a soliton \cite{ankiewicz2008dissipative}. The resulting energy redistribution makes a DS more robust and enables its energy scaling. The last becomes possible because chirp grows with energy, stretches DS, confines its peak power, and thereby suppresses a destabilization due to nonlinear effects \cite{kalashnikov2006chirped}. This regime is specific to the NGD \cite{fernandez2004chirped,Chong:06,renninger2008dissipative} and was termed dissipative soliton resonance (DSR) \cite{chang2008dissipative}.

As predicted \cite{chang2009dissipative,grelu2010dissipative}, a DSR-type energy scaling can also occur in AGD. Independent experimental and numerical studies of fiber lasers support this AGD scenario. In particular, Duan \textit{et al.} experimentally observed DSR-like rectangular pulses in an Er-doped fiber laser operating at net AGD, where the pulse duration and energy increase nearly linearly with pump, while the spectral bandwidth and peak power
remain approximately unchanged, leading to strongly chirped high-energy operation \cite{duan2011experimental}. More recently, Gene \textit{et al.} proposed and demonstrated an ultrafast AGD DS regime (U-DSAD) in which a DSR is attributed to the coaction of AGD and quintic nonlinear phase modulation, enabling high peak power and high pulse energy beyond the cubic-nonlinearity limitation \cite{Gene2023Photonix}. Together, these results provide evidence that energy-scalable, strongly chirped DS can be generated in AGD and that higher-order (quintic) nonlinearities are decisive in accessing these regimes.

However, in AGD, it competes with the usual soliton-like (zero-chirp) regime, which obeys the soliton area theorem requiring soliton peak power reduction by an extreme AGD growth or nonlinearity reduction, otherwise the energy scaling results in multi-soliton and turbulent regimes \cite{grelu2012dissipative,fermann2013ultrafast,mohamed2024energy}.

This issue of DS stability has a fundamental context and may be related to the thermodynamics of the nonlinear systems far from equilibrium. From a broader perspective, such gain-loss stabilized structures are representative of driven open many-body systems,
where detailed balance is broken by the simultaneous presence of unitary (Hamiltonian) evolution and nonunitary drive and dissipation.
This viewpoint has recently been systematized in the modern framework of driven open quantum matter and nonequilibrium universality
(see \cite{sieberer2025universality}). Such an approach was developed in photonics for the incoherent solitons of the nonlinear Schr\"{o}dinger-like equation \cite{picozzi2007towards,picozzi2009thermalization,Fischer:13,picozzi2014optical}, in particular, in connection with the possibility of soliton formation from noise in a mode-locked laser \cite{katz2006non,gordon2006self}. Recently \cite{kalashnikov2025energy}, this approach was extended to strongly chirped DS in NGD, where the entropy growth and transition to an effective negative temperature of the DS microstates limit DS energy scalability due to multi-soliton generation. It is important that the behavior of such a semi-(in-)coherent soliton mimics turbulence phenomena \cite{picozzi2014optical,nazarenko2011wave} and can be used to model it metaphorically in photonic setups \cite{longhi2009quantum}. Generalizing this approach to the AGD, which plays the role of the boson kinetic energy, could enable an analogical description of, e.g., a weakly dissipative BEC \cite{liu2024weakly,krause2025evaporative}.    

In the present work, we apply an adiabatic theory of strongly chirped DSs of the cubic-quintic CGLE to treat the NGD and AGD regimes on the same footing. Our approach yields explicit analytical criteria that define the physical existence of the two chirped solution branches and clarify when energy-scalable operation (DSR-type behavior) is possible in both dispersion regimes. On this basis, we derive closed-form spectral profiles for both NGD and AGD branches and identify a near-resonant energy regime. We show that the spectral cores (structured in AGD) carry the energy scalability via DS stretching, while the spectral cut-off translates this spectral structure into the time domain, where the first-order coherence naturally exhibits two correlation scales. A short one, a spectral-bandwidth-imposed scale, and a long one, a core-controlled scale, support a thermodynamic/microstate interpretation of the energy scalability and multi-pulse transitions in mode-locked lasers and motivate links to weakly dissipative turbulence and BEC theory.

\section{Adiabatic theory of strongly chirped DS}\label{adiabat}

Considering a large chirp as a basic condition for DS energy scaling (or a mass scaling for BEC) suggests the following assumptions \cite{kalashnikov2025energy}: 1) chirp $\psi\sim \left( \alpha/|\beta| + \kappa/\gamma \right)^{-1}\gg 1$ \cite{kalashnikov2005approaching}, which means that 2) non-dissipative linear (e.g., GDD defined by a coefficient $\beta$) and nonlinear (e.g., SPM, coefficient $\gamma$) effects prevail over the dissipative ones (e.g., spectral filtering, coefficient $\alpha$, and SAM, coefficient $\kappa$) \cite{podivilov2005heavily}. 3) As a result, a field envelope $a(t)$ evolves with a local coordinate (e.g., local time $t$ in a co-moving coordinate system for a temporal DS) slowly in comparison with the instant phase change $\phi(t)$ (\textit{adiabatic approximation}) \cite{kalashnikov2024dissipative}. Below, we follow the reasoning of \cite{kalashnikov2009chirped} and consider higher-order contributions to SPM and SAM.

Let's start with the following form of the (1+1)D cubic-quintic CGLE:

\begin{align}
\frac{\partial}{\partial z}a \! \left(z ,t \right)=-\sigma  a \! \left(z ,t \right)+\left(\alpha -\mathrm{I} \beta \right) \left(\frac{\partial^{2}}{\partial t^{2}}a \! \left(z ,t \right)\right)+
\left(\kappa +\mathrm{I} \gamma \right) P \! \left(t \right) a \! \left(z ,t \right)-\left(\kappa  \zeta +\mathrm{I} \chi \right) P \! \left(t \right)^{2} a \! \left(z ,t \right).  \label{CGLE}
\end{align}

\noindent Here $z$ is a ``longitudinal'' coordinate (e.g., propagation distance in a fiber or evolution time for BEC), $t$ is a local time (``transverse coordinate''), $a(z,t)$ is a slowly-varying field amplitude, $P(z,t)=|a(z,t)|^2$ is a power. The parameters characterizing \textit{dissipative effects} are: $\sigma$ is a saturated loss coefficient; $\alpha$ is an inverse squared bandwidth of a spectral filter (gain bandwidth in a laser); $\kappa$ is a parameter of SAM providing a loss saturation with power; $\zeta$ is a SAM saturation parameter (i.e., its dumping with $P$ for $\zeta>0$). \textit{Non-dissipative effects} are characterized by: $\beta$ is a GDD parameter which is positive for NGD and negative for AGD (the latter corresponds to a boson kinetic energy in BEC); $\gamma$ is a SPM parameter corresponding to ($\gamma>0$) self-focusing in optics or boson attraction in BEC; $\chi$ is a SPM saturation ($\chi>0$) or self-enhancement ($\chi<0$) parameter. 

We will use the traveling wave ansatz to find a solitary-pulse solution of this equation:

\begin{equation} 
    a \! \left(z ,t \right)=\sqrt{P \! \left(t \right)}\, {\mathrm e}^{\mathrm{I} \phi \left(t \right)-\mathrm{I} q z}. \label{anzatz}
\end{equation}
\noindent Here $q$ is a ``wave number''.

The substitution in Eq. (\ref{CGLE}) with the separation of real and imaginary parts gives \cite{kalashnikov2009chirped,worksheet,kalashnikov2024dissipative}

\begin{gather} \label{real}
    4 P \! \left(t \right)^{2} q -4 P \! \left(t \right)^{4} \chi -2 P \! \left(t \right) \left(\frac{d^{2}}{d t^{2}}P \! \left(t \right)\right) \beta +4 P \! \left(t \right)^{2} \left(\frac{d}{d t}\Omega \! \left(t \right)\right) \alpha +\\ \nonumber 4 P \! \left(t \right)^{3} \gamma + 
    4 P \! \left(t \right)^{2} \Omega \! \left(t \right)^{2} \beta +4 P \! \left(t \right) \left(\frac{d}{d t}P \! \left(t \right)\right) \Omega \! \left(t \right) \alpha +\left(\frac{d}{d t}P \! \left(t \right)\right)^{2} \beta =0, 
\end{gather}

\begin{gather} 
 -4 P \! \left(t \right)^{4} \zeta  \kappa -4 P \! \left(t \right)^{2} \Omega \! \left(t \right)^{2} \alpha +4 P \! \left(t \right) \left(\frac{d}{d t}P \! \left(t \right)\right) \Omega \! \left(t \right) \beta +
    4 P \! \left(t \right)^{2} \left(\frac{d}{d t}\Omega \! \left(t \right)\right) \beta + \nonumber \\ 
    4 P \! \left(t \right)^{3} \kappa -\left(\frac{d}{d t}P \! \left(t \right)\right)^{2} \alpha +2 P \! \left(t \right) \left(\frac{d^{2}}{d t^{2}}P \! \left(t \right)\right) \alpha -4 \sigma  P \! \left(t \right)^{2}=0,  \label{imag}  
\end{gather}
\noindent where $\Omega(t)\equiv d\phi(t)/dt$ is an instant frequency.

To simplify Eq.~(\ref{real}) under the assumption $\alpha \ll |\beta|$, we obtain

\begin{align} 
    \frac{P \! \left(t \right)^{2} q}{\beta}-\frac{P \! \left(t \right)^{4} \chi}{\beta}-\frac{P \! \left(t \right) \left(\frac{d^{2}}{d t^{2}}P \! \left(t \right)\right)}{2}+\frac{P \! \left(t \right)^{3} \gamma}{\beta}+P \! \left(t \right)^{2} \Omega \! \left(t \right)^{2}+\frac{\left(\frac{d}{d t}P \! \left(t \right)\right)^{2}}{4}=0. \label{assum1}
\end{align}

The next assumption is that the pulse changes slowly relative to the phase due to a large chirp. This allows the adiabatic approximation when $\frac{d^2 \sqrt{P(t)}}{dt^2}$ is negligible compared to the other terms due to the slow variation of the amplitude:

\begin{align} 
    \left(\frac{{d}^{2}}{dt^{2}}\left(\sqrt{P \! \left(t \right)}\right)\right) P \! \left(t \right)^{\frac{3}{2}}+\frac{P \! \left(t \right)^{2} q}{\beta}-\frac{P \! \left(t \right)^{4} \chi}{\beta}+\frac{P \! \left(t \right)^{3} \gamma}{\beta}+P \! \left(t \right)^{2} \Omega \! \left(t \right)^{2} = 0 \to \nonumber \\ 
    q -P \! \left(t \right)^{2} \chi +P \! \left(t \right) \gamma +\beta  \Omega \! \left(t \right)^{2}=0. \label{assum2}
\end{align}

Let's consider (\ref{imag}). After some algebra and using the adiabatic approximation, we have:

\begin{gather} 
 -P \! \left(t \right)^{2} \kappa  \zeta -\Omega \! \left(t \right)^{2} \alpha +\frac{\left(\frac{d}{d t}P \! \left(t \right)\right) \Omega \! \left(t \right) \beta}{P \! \left(t \right)}+\left(\frac{d}{d t}\Omega \! \left(t \right)\right) \beta +P \! \left(t \right) \kappa -\sigma =0.   \label{imaga}
\end{gather}

Eq. (\ref{assum2}) allows finding the instant frequency as a function of power:

\begin{equation}
    \Omega \! \left(t \right)^{2}=\frac{P \! \left(t \right)^{2} \chi}{\beta}-\frac{P \! \left(t \right) \gamma}{\beta}-\frac{q}{\beta}. \label{freq}
\end{equation}

Let's confine ourselves to the intermediate solutions with the defined limit of $\chi \to 0$. That does not mean the final solution for the DS would be physically meaningful in this limit, as shown below. The corresponding solution for $P(t)$ is:

\begin{equation} 
    P \! \left(t \right)=\frac{\gamma - \sqrt{4 \Omega \! \left(t \right)^{2} \beta  \chi +\gamma^{2}+4 \chi  q}}{2 \chi}, \label{power}
\end{equation}

\noindent so that $\lim_{\chi \to 0} P(t)=-\frac{q+\beta \Omega(t)^2}{\gamma}$.

An explicit interrelation between the instant frequency (\ref{freq}) and the power (\ref{power}) allows finding the maximum (minimum) frequency deviation 
$\Delta$ (frequency cut-off) as the limit of $P(t) \to 0$:

\begin{equation}
    \Delta^{2}=-\frac{q}{\beta}.  \label{Delta}
\end{equation} 

Excluding $P(t)$ from (\ref{imaga}) by Eqs. (\ref{power},\ref{Delta}) results in:

\begin{align} \label{eq:domega}
    \frac{d}{d t}\Omega \! \left(t \right)=\frac{\left(\Delta^{2}-\Omega \! \left(t \right)^{2}\right) \sqrt{A}\, \left(\frac{\kappa  \zeta  \left(\gamma -\sqrt{A}\right)^{2}}{4 \chi^{2}}+\Omega \! \left(t \right)^{2} \alpha -\frac{\kappa  \gamma -\kappa  \sqrt{A}-2 \sigma  \chi}{2 \chi}\right)}{\beta  \left(\left(\Delta^{2}-2 \Omega \! \left(t \right)^{2}\right) \sqrt{A}-\gamma  \Omega \! \left(t \right)^{2}\right)},\\ \nonumber
A=\gamma^{2}-4 \chi  \beta  \left(\Delta^{2}-\Omega \! \left(t \right)^{2}\right), 
\end{align}

\noindent which looks, in the limit of $\chi\to 0$, as

\begin{equation} 
    \frac{d}{d t}\Omega \! \left(t \right)= \frac{\left(\Delta^{2}-\Omega \! \left(t \right)^{2}\right)\left(\frac{\kappa \beta \left(\Delta^{2}-\Omega \left(t \right)^{2}\right) \left( \zeta \beta \left(\Delta^{2}-\Omega \left(t \right)^{2}\right)-\gamma\right)}{\gamma^{2}}+\sigma +\alpha \Omega \! \left(t \right)^{2}\right) }{\beta \left(\Delta^{2}-3 \Omega \! \left(t \right)^{2}\right)}. \label{imagal}
\end{equation}

Below, we will use the following \textit{dimensionless variables}\footnote{Note that $\tilde t$ and $\tilde\Omega$ are not Fourier-conjugate variables under the scaling~(\ref{norm}):
$\Omega t=(\gamma/\kappa)\tilde\Omega\,\tilde t$. Accordingly, the Fourier/cosine transforms are defined in the physical
pair $(t,\Omega)$ and then rewritten in $(\tilde t,\tilde\Omega)$ using the Jacobian $d\Omega=\sqrt{\gamma/(\zeta\beta)}\,d\tilde\Omega$
and the scaled spectral power $\hat S(\tilde\Omega)$.}:
\begin{equation} 
 \widetilde{\Omega}^2=\frac{|\beta|\zeta}{\gamma} \Omega^2,\qquad \widetilde{q}=\frac{\zeta}{\gamma}q,\qquad \widetilde{P}=\zeta P,\qquad \widetilde{t}=\frac{\kappa}{\sqrt{|\beta|\gamma \zeta }}t,\qquad
\widetilde{\chi}=\frac{\chi}{\gamma\zeta},\qquad
C=\frac{\alpha\gamma}{\beta\kappa},\qquad
\Sigma=\frac{\zeta}{\kappa}\sigma .   \label{norm}
\end{equation}

\noindent In practical cavity models, the saturated loss $\sigma$ (and hence $\Sigma$) is often selected self-consistently by the gain--loss balance rather than prescribed as an independent input. A short discussion of this point and its relation to master-diagram/isogain interpretations of the DS parametric space \cite{kalashnikov2025energy} is given in Appendix~A.

After rescaling, Eq. (\ref{eq:domega}) becomes

\begin{gather} \label{eq:domega2}
    \frac{d}{d \widetilde{t}}\widetilde{\Omega} \! (\widetilde{t} )=\left(\widetilde{\Delta}^{2}-\widetilde{\Omega} \! (\widetilde{t} )^{2}\right)\Upsilon\times \frac{\, \left[\widetilde{\chi}\left(2( \widetilde{\Omega} \! (\widetilde{t})^{2} C \widetilde{\chi}+\widetilde{\chi} \Sigma +   \widetilde{\Omega} \! (\widetilde{t})^{2}-   \,\widetilde{\Delta}^{2})+  \Upsilon-1\right) -\Upsilon+1\right]}{2 \left[\left(\widetilde{\Delta}^{2}-2 \widetilde{\Omega} \! (\widetilde{t} )^{2}\right) \Upsilon-\widetilde{\Omega }\! (\widetilde{t} )^{2}\right] \widetilde{\chi}^{2}},
\end{gather}
\noindent where $\Upsilon=\sqrt{1+4\widetilde{\chi}  (\widetilde{\Omega} \! (\widetilde{t})^{2}-\widetilde{\Delta}^{2})}$

The next step is to exclude the singularity in (\ref{eq:domega}): $\frac{d}{d \widetilde{t}} \widetilde{\Omega} (\widetilde{t})\neq \infty $. That means that the numerator of (\ref{eq:domega}) should become zero when the denominator tends to zero. As a result, we obtain two solutions for the cut-off frequency $\widetilde{\Delta}^2$:

\begin{equation}
\widetilde{\Delta}^2_{\mp}= \frac{\Psi^{\mp}(\widetilde{\chi})}{8(1 - C\widetilde{\chi})^2}, \label{del}
\end{equation}
where
\begin{align}
\Psi^{\mp}(\widetilde{\chi}) = -3C+6+(16\Sigma+C^{2}+2C-8)\widetilde{\chi}-16C\Sigma\widetilde{\chi}^{2} \left[\mp 3\pm (C-4)\widetilde{\chi}\right] A,
\end{align}
and
\begin{equation}
A=\sgn(\widetilde{\chi})\sqrt{(C-2)^2+16\Sigma(C\widetilde{\chi}-1)}.
\end{equation}.

The corresponding expression for the DS peak power of a solution with the regular$\widetilde{\chi} \to 0$ limit is

\begin{equation}
    \widetilde{P}_0^{\mp}= \frac{1-\sqrt{1-4\widetilde{\chi}\widetilde{\Delta}^2_{\mp}}}{2\widetilde{\chi}}. \label{pp}
\end{equation}

The signs of the DS parameters corresponding to the physical solutions are presented in Table \ref{tbl1}. Positive (negative) $\beta$ corresponds to NGD (AGD), respectively.

\paragraph{Stationary-phase spectrum (SPA).}
Define, in terms of the normalized frequency $\tilde\Omega$,
\begin{align} \label{eq:Upsilon}
\Upsilon(\tilde\Omega) &\equiv \sqrt{\,1+4\widetilde{\chi}\bigl(\tilde\Omega^{2}-\tilde\Delta^{2}\bigr)\,},\\[2mm]
G(\tilde\Omega) &\equiv \chi\!\left(2\bigl(\tilde\Omega^{2}C\widetilde{\chi}+\widetilde{\chi}\Sigma+\tilde\Omega^{2}-\tilde\Delta^{2}\bigr)+\Upsilon-1\right)-\Upsilon+1,
\end{align}
and the DS chirp from Eq.~(\ref{eq:domega2}) \footnote{We drop $\mp$ indexes in the expressions regarding NGD for convenience.}
\begin{equation}
\label{eq:omega_sweep_explicit}
\frac{d\tilde\Omega}{d\tilde t}
=\frac{\bigl(\tilde\Delta^{2}-\tilde\Omega^{2}\bigr)\,\Upsilon(\tilde\Omega)\,G(\tilde\Omega)^{2}}
{\,\Big[\bigl(\tilde\Delta^{2}-2\tilde\Omega^{2}\bigr)\,\Upsilon(\tilde\Omega)-\tilde\Omega^{2}\Big]\;\widetilde{\chi}^{2}}.
\end{equation}
The normalized power–frequency law is
\begin{equation}
\label{eq:tildeP_explicit}
\tilde P(\tilde\Omega)=\frac{1-\Upsilon(\tilde\Omega)}{2\widetilde{\chi}},
\qquad\text{with $\ \tilde\Delta^{2}$ from Eq. (\ref{del}).}
\end{equation}

Let $\tilde a(\tilde\Omega)$ denote the Fourier image of the field envelope in the normalized variables. From here, the Fourier kernel is written in dimensional $t,\Omega$. 
Under a strong chirp, the stationary-phase approximation \cite{bleistein1975asymptotic} gives\footnote{For $\gamma,\kappa$ given in $\mathrm{W}^{-1}$, the SPA prefactors $\sqrt{2\pi/(\kappa\gamma)}$ and $2\pi/\kappa\gamma$ supply the power units: $\sqrt{2\pi/(\kappa\gamma)}\sim\mathrm{W}$ so $\tilde a(\Omega)\sim\sqrt{\mathrm{W}}\cdot\mathrm{s}$ and $4\pi/(\kappa\gamma)^2\sim\mathrm{W}^2$ (the factor of 2 follows from the two equal stationary contributions with fringe averaging) so that $|\tilde a(\Omega)|^2\sim\mathrm{W}\cdot\mathrm{s}^2$. The Jacobian $d\Omega=\sqrt{\gamma/(\zeta\beta)}\,d\tilde\Omega$ converts from dimensionless $\tilde\Omega$ to physical $\Omega$, keeping Parseval’s units consistent.}

\begin{equation}
\label{eq:SPA_amplitude}
\tilde a(\tilde\Omega)\approx
\sqrt{\frac{2\pi}{\kappa\gamma}}\,
\sum_{s=\pm1}
\frac{\sqrt{P(\tilde\Omega)}}{\sqrt{\left|\frac{d\tilde\Omega}{d\tilde t}\right|_{\tilde t_s}}}
\exp\Big\{ i\big[\phi(\tilde t_s)-(\gamma/\kappa)\tilde\Omega\,\tilde t_s\big]
         + i\frac{\pi}{4}\,\mathrm{sgn}\!\left(\frac{d\tilde\Omega}{d\tilde t}\right)_{\tilde t_s}.
\end{equation}
where $\tilde t_s=\pm|\tilde t(\tilde\Omega)|$ are the stationary points determined by $\tilde\Omega(\tilde t_s)=\tilde\Omega$
(equivalently, $d\phi/d\tilde t=(\gamma/\kappa)\tilde\Omega$ at $\tilde t=\tilde t_s$),
and the dimensional frequency is $\Omega=\tilde\Omega\sqrt{\gamma/(\zeta|\beta|)}$.
The phase in \eqref{eq:SPA_amplitude} can be written purely in terms of $\tilde\Omega$:
\begin{equation}
\label{eq:SPA_phase}
\phi(\tilde t_s)-(\gamma/\kappa)\tilde\Omega\,\tilde t_s
=\frac{\gamma}{\kappa}\left[
\int_{0}^{\tilde\Omega}\frac{\tilde\Omega'}{(d\tilde\Omega/d\tilde t)(\tilde\Omega')}\,d\tilde\Omega'
- s\,\tilde\Omega\int_{0}^{\tilde\Omega}\frac{1}{(d\tilde\Omega/d\tilde t)(\tilde\Omega')}\,d\tilde\Omega'
\right],\qquad s=\pm1,
\end{equation}
with $\dfrac{d\tilde\Omega}{d\tilde t}$ substituted from \eqref{eq:omega_sweep_explicit}.

\paragraph{Spectral power (envelope).}
Averaging the $\pm$ interference in \eqref{eq:SPA_amplitude} for a symmetric single-hump pulse yields
\begin{equation}
\label{eq:SPA_power_general}
\mathcal S(\tilde\Omega)=|\tilde a(\tilde\Omega)|^2
\approx \frac{4\pi}{(\kappa\gamma)^2}\;
\frac{P(\tilde\Omega)}{\bigl|\,d\tilde\Omega/d\tilde t\,\bigr|}\,.
\end{equation}
Using \eqref{eq:tildeP_explicit} and \eqref{eq:omega_sweep_explicit} gives the fully explicit form
\begin{equation}
\label{eq:SPA_power_explicit}
\mathcal S_{\mathrm{NGD}}(\tilde\Omega)\;\approx\;
\frac{2\pi}{(\kappa\gamma)^2}\;
\tilde\chi\,\frac{(1-\Upsilon(\tilde\Omega))\,\bigl|\bigl(\tilde\Delta^{2}-2\tilde\Omega^{2}\bigr)\Upsilon(\tilde\Omega)-\tilde\Omega^{2}\bigr|}
{\bigl(\tilde\Delta^{2}-\tilde\Omega^{2}\bigr)\,\Upsilon(\tilde\Omega)\,\bigl|G(\tilde\Omega)\bigr|^{2}}.
\end{equation}
which depends only on $\tilde\Omega$, $\widetilde{\chi}$, $C$, $\Sigma$, and $\tilde\Delta^{2}$. Near the spectral cut-off $|\tilde\Omega|=\tilde\Delta$, the two stationary points coalesce and the ordinary SPA loses uniformity. \footnote{One may use the CFU--Airy uniformization (Appendix~B, Eq.~\eqref{eq:airy-min} with $S_{\mathrm{edge}}\equiv\lim_{\tilde{\Omega}\to\tilde{\Delta}^{-}}S_{\mathrm{NGD}}(\tilde{\Omega})$)
to remove the $0/0$ singularity and obtain a finite edge value with an exponentially small roll-off.}

\paragraph{Interference fringes and multi-horn spectra (beyond fringe averaging).}
In the coherent two-saddle SPA, the spectral field is a \emph{sum} of two stationary-point contributions. Their relative phase produces oscillatory
spectral structure (fringes) on top of a smooth envelope. This mechanism is generic for both NGD and AGD. The difference is that, in NGD, it manifests itself across the entire interval defined by the spectral cut-off, whereas in AGD it mainly affects the central core and becomes weaker in the far wings.

Eq.~(\ref{eq:SPA_amplitude}) gives the coherent SPA field as
\begin{equation}\label{eq:27}
\tilde a(\Omega)\approx \sum_{s=\pm 1} A_s(\Omega)\,e^{i\Phi_s(\Omega)},
\qquad
A_s(\Omega)=\sqrt{\frac{2\pi}{\kappa\gamma}}\,
\frac{\sqrt{\tilde P(\Omega)}}{\sqrt{\Bigl|\frac{d\tilde\Omega}{d\tilde t}\Bigr|_{\tilde t_s}}},
\end{equation}
where $\Phi_s(\Omega)$ denotes the total saddle phase, i.e., the full exponent of the $s$-th term in Eq.~(23).
Its main contribution $\phi(\tilde t_s)-(\gamma/\kappa)\tilde\Omega\,\tilde t_s$ is given explicitly by Eq.~(24),
while the remaining $\pm\pi/4$ term is the standard stationary-phase signature.
The corresponding spectral power contains an interference term,
\begin{equation}
\tilde{\mathcal S}(\Omega)=|\tilde a(\Omega)|^2
= \sum_{s=\pm} |A_s(\Omega)|^2 + 2|A_+(\Omega)A_-(\Omega)|\cos\Delta\Phi(\Omega),
\qquad
\Delta\Phi\equiv \Phi_+ - \Phi_- .
\label{eq:fringed-spectrum}
\end{equation}
For a symmetric pulse $|A_+|\simeq|A_-|$, so one can write
\begin{equation}
\tilde{\mathcal S}(\Omega)\simeq \tilde{\mathcal S}_{\rm env}(\Omega)\,
\Bigl[1+\cos\Delta\Phi(\Omega)\Bigr],
\qquad
\tilde{\mathcal S}_{\rm env}(\Omega)\equiv \frac{4\pi}{(\kappa\gamma)^2}\,
\frac{\tilde P(\Omega)}{\Bigl|\frac{d\tilde\Omega}{d\tilde t}\Bigr|},
\label{eq:env-times-fringe}
\end{equation}
where $\tilde{\mathcal S}_{\rm env}$ coincides with the fringe-averaged envelope used in Eq.~(\ref{eq:SPA_power_general}).

Thus, the fringe-averaging step leading to Eq.~(\ref{eq:SPA_power_general}) is equivalent to replacing
$\bigl[1+\cos\Delta\Phi(\Omega)\bigr]\mapsto 1$. However, in fully coherent simulations (deterministic phase and sufficient spectral resolution),
the cosine term survives and produces a small number of pronounced maxima near the spectral center together with weaker oscillations in the wings.

A closely related effect was observed experimentally in NGD chirped-pulse oscillators. For example, the Cr:ZnS spectra in Fig.~2 of
Ref.~\cite{kalashnikov2025energy} show a central ``finger'' accompanied by a slow (low-frequency) modulation on the wing, which is qualitatively
consistent with incomplete fringe averaging in Eq.~(\ref{eq:env-times-fringe}).%
\footnote{Narrow, high-frequency modulations may have extrinsic origins (e.g., intracavity absorption or measurement artifacts) and should not be
conflated with the intrinsic two-saddle interference fringes (see, e.g., Ref.~\cite{Sorokin:23}).}

The fringe phase can be expressed through the chirp law $\frac{d\tilde\Omega}{d\tilde t}$ in Eq.~(\ref{eq:omega_sweep_explicit}):
\begin{equation}
\tilde t(\tilde\Omega)=\int_{0}^{\tilde\Omega}\frac{d\tilde\Omega'}{\left(d\tilde\Omega/d\tilde t\right)(\tilde\Omega')},
\qquad
\Delta\Phi(\Omega)=2(\gamma/\kappa)\tilde\Omega\,\tilde t(\tilde\Omega)+\Delta\Phi_0,
\label{eq:phase-diff}
\end{equation}
so fringe maxima satisfy $\Delta\Phi(\Omega)=2\pi m$ ($m\in\mathbb Z$), while minima satisfy $\Delta\Phi(\Omega)=(2m+1)\pi$.
Accordingly, the ``multi-horn'' structure is the coherent $\pm$-saddle interference implied by Eq.~(\ref{eq:27}), superimposed on the smooth
envelope $\tilde{\mathcal S}_{\rm env}$.\\

As shown below in Eqs.~\eqref{eq:NGD-Lorentz}–\eqref{eq:NGD-Lorentz-E}, the NGD spectrum admits a node-regularized truncated-Lorentz form consistent with the NGD simulations and experiments. Finite but small $|\chi|$ only produces a slight rounding of the cutoff (see below). Near the spectral cut-off $|\tilde\Omega|=\tilde\Delta$, where the two stationary points coalesce, a uniform Airy treatment (Appendix~B) removes the $0/0$-singularity and yields a finite edge value with an exponentially small roll-off.

Since $\tilde P(\tilde\Omega)= (1-\Upsilon)/(2\tilde\chi)$, one has $\tilde P\ge 0 \Leftrightarrow \Upsilon\le 1$ for $\tilde\chi>0$ and $\tilde P\ge 0 \Leftrightarrow \Upsilon\ge 1$ for $\tilde\chi<0$.
In both cases, this implies $|\tilde\Omega|\le\tilde\Delta$. Thus, the NGD spectrum is \emph{always} truncated at
$|\tilde\Omega|=\tilde\Delta$, and regions with $|\tilde\Omega|>\tilde\Delta$ are unphysical even if the radicand in $\Upsilon$ is positive.

\textit{For AGD} with $C<0$ and $\tilde\Delta^2<0$, one has
\begin{equation}
\label{eq:SPA_power_explicit2}
\mathcal S_{\mathrm{AGD}}(\tilde\Omega)\;\approx\;
\frac{2\pi}{(\kappa\gamma)^2}\;
\tilde\chi\,\frac{(1+\Upsilon(\tilde\Omega))\,\bigl|\bigl(\tilde\Delta_{-}^{2}-2\tilde\Omega^{2}\bigr)\Upsilon(\tilde\Omega)-\tilde\Omega^{2}\bigr|}
{\bigl(\tilde\Delta_{-}^{2}-\tilde\Omega^{2}\bigr)\,\Upsilon(\tilde\Omega)\,\bigl|G(\tilde\Omega)\bigr|^{2}}.
\end{equation}

\noindent where only $\tilde\Delta_{-}$-branch, $\widetilde{\chi}>0$ persist, and (see below)

\begin{equation}
\label{eq:tildeP_explicit2}
\tilde P(\tilde\Omega)=\frac{1+\Upsilon(\tilde\Omega)}{2\widetilde{\chi}}.
\end{equation}

Taking into account our normalization of time (\ref{norm}), we can use the scaled spectral power $\hat{S}\left( \tilde{\Omega} \right)=\frac{\gamma^2 \zeta \left| \beta \right|}{2\pi \kappa}S\left( \tilde{\Omega} \right)$.

Now we apply the \textit{regularization procedure} to exclude the node from the expression for the spectrum profile. 

Let $\; D\equiv\tilde\Delta^{2}>0,\quad u=\tilde\Omega/\sqrt{D}\in[0,1],\quad
Y(u)=1-3u^{2},\quad X(u)=1-u^{2}.
\;$ In the \textit{NGD limit} $\tilde\chi \to 0$, the SPA envelope is
\begin{equation}\label{eq:s0ngd}
  \mathcal S_{\text{NGD}}^{(0)}(\tilde\Omega)=\frac{4\pi}{(\kappa\gamma)^2}\,
\frac{\big|D-3\tilde\Omega^{2}\big|}{\,D X(u)^{2}+(C-1)X(u)+\Sigma\,},\qquad X(u)=1-u^{2}.  
\end{equation}

Introduce the quadratic denominator in the \(Y\)-basis (about the node \(Y=0\), i.e.\ \(u=1/\sqrt{3}\)):
\[
\mathcal D(Y)\equiv D X^{2}+(C-1)X+\Sigma
=A_{2}Y^{2}+A_{1}Y+A_{0},
\]
\vspace{-1.25ex}
\[
A_{2}=\frac{D}{9},\qquad
A_{1}=\frac{4D}{9}+\frac{C-1}{3},\qquad
A_{0}=\frac{4D}{9}+\frac{2(C-1)}{3}+\Sigma.
\]

In the node window, linearizing the denominator in Eq.~(\ref{eq:s0ngd}) as $A_0+A_1Y$ (i.e., setting $A_2\!\to0$) reproduces the truncated Lorentzian
form \eqref{eq:NGD-Lorentz} with the width \eqref{eq:GammaL}. Keeping $A_2\neq0$ yields the next-order correction beyond Lorentzian and
manifests mainly as a mild smoothing of the cutoff at $|\tilde\Omega|=\tilde\Delta$. 

Expanding \(1/\mathcal D(Y)\) at \(Y=0\) up to \(Y^{2}\) gives
\[
\frac{1}{\mathcal D(Y)}=\frac{1}{A_0}
-\frac{A_1}{A_0^{2}}\,Y
+\frac{A_1^{2}-A_0A_2}{A_0^{3}}\,Y^{2}+O(Y^{3}).
\]

Inserting the second-order expansion of $1/\mathcal{D}(Y)$ around $Y=0$ into Eq.~(\ref{eq:s0ngd}) yields an explicit NGD spectrum
that captures the nodal behavior analytically:
\begin{equation}
\hat S_{\text{NGD}}^{(0)}(\tilde\Omega)
=\frac{\gamma^{2}\zeta|\beta|}{2\pi\kappa}\,\mathcal S_{\text{NGD}}^{(0)}(\tilde\Omega)
\;\approx\;
\frac{2\,\zeta|\beta|}{\kappa^{3}}\;D\;\Bigg[
\frac{\lvert Y\rvert}{A_{0}}
-\frac{A_{1}}{A_{0}^{2}}\,\lvert Y\rvert\,Y
+\frac{A_{1}^{2}-A_{0}A_{2}}{A_{0}^{3}}\,\lvert Y\rvert\,Y^{2}
\Bigg],\quad Y=1-3u^{2}.
\end{equation}
Note that the spectrum still vanishes at the node ($Y=0$). The regularization here refers to replacing the denominator with a local
analytic approximation, which makes the subsequent energy integral explicit.

\noindent With the energy definition \(E=\int_{-\sqrt{D}}^{\sqrt{D}}\hat S\,d\tilde\Omega/(2\pi)\), the evenness gives
\[
E=\frac{1}{\pi}\int_{0}^{\sqrt{D}}\hat S(\tilde\Omega)\,d\tilde\Omega
=\frac{\sqrt{D}}{\pi}\int_{0}^{1}\hat S(\sqrt{D}\,u)\,du.
\]
Using the three elementary integrals
\[
\mathcal I_{0}=\!\int_{0}^{1}\!\lvert Y\rvert\,du=\frac{4\sqrt{3}}{9},\quad
\mathcal I_{1}=\!\int_{0}^{1}\!\lvert Y\rvert\,Y\,du=-\frac{4}{5}+\frac{16\sqrt{3}}{45},\quad
\mathcal I_{2}=\!\int_{0}^{1}\!\lvert Y\rvert\,Y^{2}\,du=\frac{16}{35}+\frac{32\sqrt{3}}{105},
\]
we obtain the closed–form energy (node–regularized NGD)\footnote{Our node–regularized NGD energy,

\(
E_{\!{\rm NGD}}^{(0)}=\frac{2\,\zeta|\beta|}{\pi\kappa^{3}}\,D\sqrt{D}\Big[\frac{\mathcal I_{0}}{A_{0}}
-\frac{A_{1}}{A_{0}^{2}}\mathcal I_{1}
+\frac{A_{1}^{2}-A_{0}A_{2}}{A_{0}^{3}}\mathcal I_{2}\Big]
\),
reduces to the truncated Lorentzian result of \cite{podivilov2005heavily} if one linearizes the denominator at the node
(\(Y=0\)) by setting \(A_{2}\!\to 0\). In that case,
\(\hat S(\tilde\Omega)\propto |Y|/(A_{0}+A_{1}Y)\)
maps (via \(Y\propto \tilde\Omega-\tilde\Omega_{0}\)) to
\(\hat S(\tilde\Omega)\simeq \mathcal A/[\Gamma^{2}+(\tilde\Omega-\tilde\Omega_{0})^{2}]\)
with
\(\Gamma=(\sqrt{D}/\sqrt{3})\,A_{0}/|A_{1}|\),
so the energy is \(\propto \arctan(\sqrt{D}/\Gamma)\), exactly as in \cite{podivilov2005heavily}.
Keeping \(A_{2}\neq 0\) yields the next–order correction beyond the Lorentzian one (the
\(\mathcal I_{2}\) term above).}:
\begin{equation}\label{eq:energy1}
E_{\text{NGD}}^{(0)}
\;\approx\;
\frac{2\,\zeta|\beta|}{\pi\,\kappa^{3}}\;D\sqrt{D}\;
\Bigg[
\frac{\mathcal I_{0}}{A_{0}}
-\frac{A_{1}}{A_{0}^{2}}\;\mathcal I_{1}
+\frac{A_{1}^{2}-A_{0}A_{2}}{A_{0}^{3}}\;\mathcal I_{2}
\Bigg].
\end{equation}

Using \(\Upsilon(\tilde\Omega)\) and \(G(\tilde\Omega)\) from (\ref{eq:Upsilon})–(\ref{eq:omega_sweep_explicit}), let define
\[
Y_{\chi}(\tilde\Omega)\equiv(\tilde\Delta^{2}-2\tilde\Omega^{2})\,\Upsilon(\tilde\Omega)-\tilde\Omega^{2},
\qquad
\Xi_{\chi}(\tilde\Omega)\equiv
\frac{(\tilde\Delta^{2}-\tilde\Omega^{2})\,\Upsilon(\tilde\Omega)\,\bigl|G(\tilde\Omega)\bigr|^{2}}
{\tilde\chi\,\bigl(1-\Upsilon(\tilde\Omega)\bigr)}.
\]
Then, the SPA envelope reads
\[
S(\tilde\Omega)=\frac{2\pi}{(\kappa\gamma)^{2}}\,
\frac{\bigl|Y_{\chi}(\tilde\Omega)\bigr|}{\Xi_{\chi}(\tilde\Omega)}\,.
\]

One may expand \(\Xi_{\chi}\) about \(\Om_{0}\) and express the result as a quadratic in \(Y_{\chi}\):
\[
\Xi_{\chi}(\Om)\;\approx\;A_{0}(\tilde\chi)+A_{1}(\tilde\chi)\,Y_{\chi}(\Om)+A_{2}(\tilde\chi)\,Y_{\chi}(\Om)^{2},
\]
with the \emph{explicit} coefficient formulas
\[
A_{0}=\Xi_{\chi}(\tilde\Omega_{0}),\qquad
A_{1}=\frac{\Xi'_{\chi}(\tilde\Omega_{0})}{Y'_{\chi}(\tilde\Omega_{0})},\qquad
A_{2}=\frac{\Xi''_{\chi}(\tilde\Omega_{0})-A_{1}\,Y''_{\chi}(\tilde\Omega_{0})}{2\,[Y'_{\chi}(\tilde\Omega_{0})]^{2}},
\]
and the derivatives
\[
\Ups'(\Om)=\frac{4\tilde\chi\,\Om}{\Ups},\quad
\Ups''(\Om)=\frac{4\tilde\chi}{\Ups}-\frac{16\tilde\chi^{2}\Om^{2}}{\Ups^{3}},\quad
Y_{\chi}'(\Om)=-4\Om\,\Ups+(\Del^{2}-2\Om^{2})\Ups'(\Om)-2\Om,
\]
\[
Y_{\chi}''(\Om)=-4\Ups-4\Om\,\Ups'
+(\Del^{2}-2\Om^{2})\Ups''(\Om)-4\Om\,\Ups' -2.
\]

Let us introduce a small \(\chi\)-vanishing regulator \(\delta_{\chi}=K\,\tilde\chi^{2}\) (e.g. \(K=\tfrac{4}{3}\Del^{2}\)),
and define the node–regularized spectrum
\begin{equation}\label{eq:NGDspa1}
  \hat S_{\mathrm{NGD}}(\Om;\tilde\chi)=
\frac{\gamma^{2}\zeta|\beta|}{2\pi\kappa}\,S_{\mathrm{NGD}}(\Om;\tilde\chi)
=\frac{2\,\zeta|\beta|}{\kappa^{3}}\;
\frac{\sqrt{\,Y_{\chi}(\Om)^{2}+\delta_{\chi}^{2}\,}}{\;
\big|\,A_{0}+A_{1}\,Y_{\chi}(\Om)+A_{2}\,Y_{\chi}(\Om)^{2}\,\big|}\,.
\end{equation}

This reduces to the \(\chi\!\to 0\) node–regularized formula and
captures the \(\chi\)-dependent node shift through \(\Om_{0}(\chi)\).

With \(E=\int_{-\Del}^{\Del}\hat S\,d\Om/(2\pi)\) and evenness, we have
\begin{equation}\label{eq:NGDen1}
E_{\mathrm{NGD}}(\tilde\chi)=\frac{1}{\pi}\int_{0}^{\Del}\hat S_{\mathrm{NGD}}(\Om;\tilde\chi)\,d\Om
=\frac{2\,\zeta|\beta|}{\pi\,\kappa^{3}}
\int_{0}^{\Del}\!
\frac{\sqrt{\,Y_{\chi}(\Om)^{2}+\delta_{\chi}^{2}\,}}{\;
\big|\,A_{0}+A_{1}\,Y_{\chi}(\Om)+A_{2}\,Y_{\chi}(\Om)^{2}\,\big|}\,d\Om\;.
\end{equation}

Let \(\theta\equiv Y'_{\chi}(\tilde\Omega_{0})\) and set \(y:=Y_{\chi}(\tilde\Omega)\).
Using the expansion coefficients \(A_{0,1,2}\) defined above, the node–regularized energy takes the compact form
\[
E_{\mathrm{NGD}}(\chi)
=\frac{1}{\pi|\theta|}\int_{y_{L}}^{y_{U}}
\frac{\sqrt{\,y^{2}+\delta_{\chi}^{2}\,}}{\;\bigl|\,A_{0}+A_{1}y+A_{2}y^{2}\,\bigr|}\,dy,
\]
with \(y_{L}=Y_{\chi}(0)\) and \(y_{U}=Y_{\chi}(\tilde\Delta)\).

\paragraph{(i) Linearized denominator (\(A_{2}=0\)).}
If one keeps only the linear term in the denominator expansion, then
\begin{equation} \label{eq:NGDen2}
   E_{\mathrm{NGD}}(\tilde\chi)\;\simeq\;\frac{1}{\pi|\theta|}\int_{y_L}^{y_U}
\frac{\sqrt{y^{2}+\delta_{\chi}^{2}}}{\,|A_{0}+A_{1}y|}\,dy
\;=\;\frac{2\,\zeta|\beta|}{\pi\,\kappa^{3}\,|\theta|\,|A_{1}|}\;
\Biggl[
\sqrt{y^{2}+\delta_{\chi}^{2}}\,
\arctan\!\frac{|A_{1}|\,y}{A_{0}\sqrt{y^{2}+\delta_{\chi}^{2}}}
\Biggr]_{y_L}^{y_U}. 
\end{equation}

In the $\chi=0$ NGD limit, the node is removed, and the observable spectrum takes the central, node-free truncated–Lorentz form \cite{podivilov2005heavily}:
\begin{equation}
\label{eq:NGD-Lorentz}
\hat S_{\rm NGD}^{(0)}(\tilde\Omega)\;=\;
\frac{\mathcal A_L \Gamma_L^{2}} {\Gamma_L^{2}+\tilde\Omega^{2}}\,
\Theta\!\bigl(\tilde\Delta^{2}-\tilde\Omega^{2}\bigr).
\end{equation}
Here $\Theta$ denotes the Heaviside step function, and $\mathcal A_L$ is the SPA normalization fixed by the prefactors in Eqs.~(\ref{eq:SPA_amplitude})--(\ref{eq:SPA_power_explicit}). The node-window width is
\begin{equation}
\label{eq:GammaL}
\Gamma_L=\frac{\tilde\Delta}{\sqrt{3}}\;\frac{A_0}{|A_1|},
\end{equation}
with $A_0,A_1$ evaluated at $\chi=0$ (see the definitions around Eq.~(\ref{eq:s0ngd})). The corresponding energy is
\begin{equation}
\label{eq:NGD-Lorentz-E}
E_{\rm NGD}^{(0)}=\frac{\mathcal A_L}{\pi}\arctan\!\Big(\frac{\tilde\Delta}{\Gamma_L}\Big).
\end{equation}

Returning to the NGD node regularization, we note that the $\tilde\chi\to 0$ NGD energy coincides with the result of \cite{podivilov2005heavily}. The apparent zero of the raw SPA envelope at $\tilde\Omega_0=\tilde\Delta/\sqrt{3}$ is a removable artifact of the $0/0$ structure in the node window.
Linearizing the denominator in the $Y=1-3u^2$ basis and eliminating this removable singularity yields the node-free truncated--Lorentzian profile~\eqref{eq:NGD-Lorentz} with the width $\Gamma_L$ given by Eq.~\eqref{eq:GammaL} and energy given by Eq.~\eqref{eq:NGD-Lorentz-E}, in qualitative agreement with the NGD simulations and experiments discussed in \cite{kalashnikov2025energy}.

\paragraph{(ii) Quadratic denominator kept (\(A_{2}\neq 0\)): elementary closed form.}
Let retain the quadratic term but keep the linearization \(y=Y_{\chi}(\Om)\approx \theta (\Om-\Om_{0})\).
Set \(y=\delta_{\chi}\sinh \tau\) (so \(\sqrt{y^{2}+\delta_{\chi}^{2}}=\delta_{\chi}\cosh \tau\)), and define
\[
\mathfrak{A}=A_{2}\,\delta_{\chi}^{2},\qquad \mathfrak{B}=A_{1}\,\delta_{\chi},\qquad \mathfrak{G}=A_{0},\qquad
\mathfrak{D}=4\mathfrak{A} \mathfrak{G}-\mathfrak{B}^{2}.
\]
Then
\begin{equation} \label{eq:NGDen3}
   E_{\mathrm{NGD}}(\tilde\chi)\;\simeq\;\frac{2\,\zeta|\beta|}{\pi\,\kappa^{3}\,|\theta|}\,
\int_{\tau_L}^{\tau_U}\frac{\delta_{\chi}^{2}\cosh^{2}\tau}{\,\mathfrak{A}\sinh^{2}\tau+\mathfrak{B}\sinh \tau+\mathfrak{G}\,}\,d\tau,
\quad (\tau_{L,U}=\operatorname{arcsinh}(y_{L,U}/\delta_{\chi})). 
\end{equation}

\noindent This integral evaluates in elementary functions. The final form depends only on the sign of the
discriminant. Let define
\[
W(y)=\sqrt{y^{2}+\delta_{\chi}^{2}},
\]
then
\begin{equation} \label{eq:NGDen4}
E_{\mathrm{NGD}}(\tilde\chi)=\frac{2\,\zeta|\beta|}{\pi\,\kappa^{3}\,|\theta|}
\begin{cases}
\bigl[\Phi_{\arctan}(y)\bigr]_{y_L}^{y_U}, & \mathfrak{D}>0,\\[0.25em]
\bigl[\Phi_{\operatorname{artanh}}(y)\bigr]_{y_L}^{y_U}, & \mathfrak{D}<0,
\end{cases}    
\end{equation}

\noindent with
\[
\Phi_{\arctan}(y)=
\frac{1}{2\mathfrak{A}}\ln(\mathfrak{A}y^{2}+\mathfrak{B}y+\mathfrak{G})
+\frac{\mathfrak{B}}{2\mathfrak{A}\sqrt{\mathfrak{D}}}\,\operatorname{asinh}\!\Big(\frac{y}{\delta_{\chi}}\Big)
-\frac{1}{\sqrt{\mathfrak{D}}}\,\arctan\!\Big(\frac{2\mathfrak{A}\,W(y)+\mathfrak{B}\,y}{\sqrt{\mathfrak{D}}}\Big),
\quad (\mathfrak{D}>0),
\]
\[
\Phi_{\operatorname{artanh}}(y)=
\frac{1}{2\mathfrak{A}}\ln(\mathfrak{A}y^{2}+\mathfrak{B}y+\mathfrak{G})
+\frac{\mathfrak{B}}{2\mathfrak{A}\sqrt{-\mathfrak{D}}}\,\operatorname{asinh}\!\Big(\frac{y}{\delta_{\chi}}\Big)
-\frac{1}{\sqrt{-\mathfrak{D}}}\,\operatorname{artanh}\!\Big(\frac{2\mathfrak{A}\,W(y)+\mathfrak{B}\,y}{\sqrt{-\mathfrak{D}}}\Big),
\quad (\mathfrak{D}<0).
\]
If \(\mathfrak{D}=0\), the equations follow by a continuous limit.  
That reduces to the \(\arctan\)–only result above when \(A_{2}\to 0\).

Now, let's consider \textbf{AGD} and return to Eq. (\ref{eq:SPA_power_explicit2}):

\begin{equation} \label{eq:AGDspa1}
\hat S_{\mathrm{AGD}}(\tilde\Omega)\;=\;\frac{\gamma^{2}\zeta|\beta|}{2\pi\kappa}\,
\mathcal S_{\mathrm{AGD}}(\tilde\Omega)
=\frac{2\,\zeta|\beta|}{\kappa^{3}}\,
\chi\,
\frac{\bigl|\bigl(\tilde\Delta_{-}^{2}-2\tilde\Omega^{2}\bigr)\,\Upsilon(\tilde\Omega)-\tilde\Omega^{2}\bigr|\,
\bigl(1+\Upsilon(\tilde\Omega)\bigr)}
{\bigl(\tilde\Delta_{-}^{2}-\tilde\Omega^{2}\bigr)\,\Upsilon(\tilde\Omega)\,\bigl|G(\tilde\Omega)\bigr|^{2}}\,.
\end{equation}

\noindent Since
\[
\Upsilon(\tilde\Omega)=\sqrt{\,1+4\tilde \chi\bigl(\tilde\Omega^{2}-\tilde\Delta_{-}^{2}\bigr)} > 1
\quad(\forall\,\tilde\Omega,\ \tilde \chi>0),
\]
we have \(P=(1+\Upsilon)/(2\tilde \chi)>0\) for all \(\tilde\Omega\), and the spectrum has no truncation. Its asymptotic tail is
\begin{equation} \label{eq:AGDspa2}
\mathcal S_{\mathrm{AGD}}(\tilde\Omega)\sim
\frac{2\pi}{(\kappa\gamma)^2}\,
\frac{1}{(1+C\tilde\chi)^{2}\,\sqrt{\tilde\chi}\,|\tilde\Omega|^{3}}\quad(|\tilde\Omega|\to\infty).
\end{equation}

\noindent so the energy is finite. $\tilde\Delta$ serves as a scale parameter in AGD, controlling only the shape/level of the spectrum.

Although in AGD (\(\beta<0,\ C<0,\ \tilde \chi>0\)) there is no finite spectral cut-off, the DS remains \emph{strongly chirped}. Indeed, one has from (\ref{eq:domega2}) for large \(|\tilde\Omega|\):
\[
\frac{d\tilde\Omega}{d\tilde t}\;\sim\;2\,(1+C\tilde \chi)^{2}\,\tilde\Omega^{4},
\]
so the phase curvature \(\phi''=d\tilde\Omega/d\tilde t\) is large and monotone. 

One has to note that there is no spectral node in AGD. Writing $\tilde\Delta_{-}^{2}=-D_{a}$ with $D_{a}>0$ and $\Upsilon(\tilde\Omega)=\sqrt{1+4\tilde \chi(\tilde\Omega^{2}+D_{a})}>1$, we have
$Y_{\chi}(\tilde\Omega)=(\tilde\Delta_{-}^{2}-2\tilde\Omega^{2})\Upsilon(\tilde\Omega)-\tilde\Omega^{2}
= -\big[(D_{a}+2\tilde\Omega^{2})\Upsilon(\tilde\Omega)+\tilde\Omega^{2}\big]<0$ for all $\tilde\Omega$.
Hence the numerator of $\mathcal S_{\mathrm{AGD}}$ never vanishes and the node-regularization used in NGD is not required in AGD.

The expression for the DS energy in AGD is:
\begin{equation} \label{eq:enAGD}
 E_{\mathrm{AGD}}=\frac{1}{\pi}\int_{0}^{\infty}\hat S_{\mathrm{AGD}}(\tilde\Omega)\,d\tilde\Omega
=\frac{2\,\zeta|\beta|}{\pi\,\kappa^{3}}\int_{0}^{\infty}
\chi\,
\frac{\bigl|\bigl(\tilde\Delta_{-}^{2}-2\tilde\Omega^{2}\bigr)\,\Upsilon(\tilde\Omega)-\tilde\Omega^{2}\bigr|\,
\bigl(1+\Upsilon(\tilde\Omega)\bigr)}
{\bigl(\tilde\Delta_{-}^{2}-\tilde\Omega^{2}\bigr)\,\Upsilon(\tilde\Omega)\,\bigl|G(\tilde\Omega)\bigr|^{2}}
\,d\tilde\Omega\,,
\end{equation}
\noindent which is finite due to the \( |\tilde\Omega|^{-3} \) tail.

\noindent Here
\[
\Upsilon(\tilde\Omega)=\sqrt{1+4\tilde\chi\bigl(\tilde\Omega^{2}-\tilde\Delta_{-}^{2}\bigr)}\ (>1),\qquad
G(\tilde\Omega)=\tilde \chi\!\left(2\bigl(\tilde\Omega^{2}C\tilde \chi+\tilde \chi\Sigma+\tilde\Omega^{2}-\tilde\Delta_{-}^{2}\bigr)+\Upsilon-1\right)-\Upsilon+1,
\]
and \(\hat S_{\mathrm{AGD}}=\frac{\gamma^{2}\zeta|\beta|}{2\pi\kappa}\,\mathcal S_{\mathrm{AGD}}\) with \(\mathcal S_{\mathrm{AGD}}\) given by (\ref{eq:AGDspa1}).

Let \(D_a \equiv |\tilde\Delta_{-}^{2}|>0\) and \(\mathfrak b \equiv 1+4\tilde\chi D_a>1\).
With the radical–free substitution
\begin{equation} \label{eq:AGD_subst}
   \Upsilon=\sqrt{\mathfrak b}\,\cosh\tau,\qquad
\tilde\Omega^{2}=\frac{\mathfrak b}{4\tilde \chi}\,\sinh^{2}\tau,\qquad
d\tilde\Omega=\frac{\sqrt{\mathfrak b}}{2\sqrt{\tilde \chi}}\cosh\tau\,d\tau, 
\end{equation}
\noindent we have the exact identity
\[
\tilde \Delta_{-}^{2}-\tilde\Omega^{2} = \frac{1-\Upsilon^{2}}{4\tilde\chi}
= -\frac{\mathfrak b\cosh^{2}\tau-1}{4\tilde\chi}.
\]
Consequently,
\[
\bigl|(\tilde\Delta_{-}^{2}-2\tilde\Omega^{2})\,\Upsilon-\tilde\Omega^{2}\bigr|
= -\Bigl[(D_a+2\tilde\Omega^{2})\,\Upsilon+\tilde\Omega^{2}\Bigr],
\]
and the integrand of (\ref{eq:enAGD}) becomes a rational function of \(\cosh\tau\).

Let write \(G\) as a quadratic in \(\Upsilon\) and then in \(\cosh\tau\):
\[
G(\tilde\Omega)=p_{2}\,\Upsilon^{2}+p_{1}\,\Upsilon+p_{0},\qquad
p_{2}=\frac{1+\tilde\chi C}{2},\quad p_{1}=\tilde\chi-1,\quad p_{0}=\frac{1}{2}-\tilde\chi-\frac{\tilde\chi C}{2}\,\mathfrak{b}+2\tilde\chi^{2}\Sigma,
\]
\[
\tilde L(\tau)\equiv \Lambda_{2}\cosh^{2}\tau+\Lambda_{1}\cosh\tau+\Lambda_{0},\qquad
\Lambda_{2}=p_{2}\,\mathfrak{b},\ \ \Lambda_{1}=p_{1}\sqrt{\mathfrak{b}},\ \ \Lambda_{0}=p_{0},
\]
and denote its discriminant by
\[
\mathfrak D_{\Lambda} \equiv 4\,\Lambda_{2}\Lambda_{0}-\Lambda_{1}^{2}.
\]

Using \eqref{eq:AGD_subst}, the integrand of \eqref{eq:enAGD} becomes a rational function of \(\cosh\tau\):
\begin{equation}  \label{eq:enAGD2}
  E_{\mathrm{AGD}}
=\frac{2\,\zeta|\beta|}{\pi\,\kappa^{3}}\,
\Bigl[\Psi(\tau)\Bigr]_{\tau=0}^{\tau=\infty},
\qquad
\Psi'(\tau)=
\frac{\mathcal{R}\!\bigl(\cosh\tau\bigr)}{\bigl(\mathfrak{b}\cosh^{2}\tau-1\bigr)\,\tilde L(\tau)^{2}},  
\end{equation}

\noindent where \(\mathcal{R}\) is a cubic polynomial in \(\cosh\tau\) whose coefficients are algebraic in \((\tilde\chi,C,\Sigma,D_{\!a})\).
Therefore \(\Psi(\tau)\) has an elementary closed form.

Then
\begin{equation}  \label{eq:enAGD3}
 E_{\mathrm{AGD}}=\frac{2\,\zeta|\beta|}{\pi\,\kappa^{3}}\,
\bigl[\Psi(\tau)\bigr]_{\tau=0}^{\infty},\qquad
\Psi(\tau)=\Phi_{\text{log}}(\tau)+
\begin{cases}
\displaystyle \frac{\mathcal A_{1}}{\sqrt{\mathfrak{D}_{\Lambda}}}\,
\arctan\!\Bigl(\frac{2\Lambda_{2}\sinh\tau+\Lambda_{1}}{\sqrt{\mathfrak{D}_{\Lambda}}}\Bigr),
& \mathfrak{D}_{\Lambda}>0,\\[1.25ex]
\displaystyle \frac{\mathcal A_{1}}{\sqrt{-\mathfrak{D}_{\Lambda}}}\,
\operatorname{artanh}\!\Bigl(\frac{2\Lambda_{2}\sinh\tau+\Lambda_{1}}{\sqrt{-\mathfrak{D}_{\Lambda}}}\Bigr),
& \mathfrak{D}_{\Lambda}<0,\\[1.25ex]
\displaystyle \mathcal A_{1}\,\frac{2\Lambda_{2}\sinh\tau+\Lambda_{1}}
{\,\Lambda_{2}\sinh^{2}\tau+\Lambda_{1}\sinh\tau+\Lambda_{0}\,},
& \mathfrak{D}_{\Lambda}=0,
\end{cases}   
\end{equation}

\noindent where the logarithmic part is
\[
\Phi_{\text{log}}(\tau)=
\frac{\mathcal A_{0}}{2\Lambda_{2}}\ln\!\tilde L(\tau)\;
-\;\frac{\mathcal B_{0}}{2\mathfrak b}\ln\!\bigl(\mathfrak b\cosh^{2}\tau-1\bigr)
\;+\;\frac{\mathcal B_{1}}{\sqrt{\mathfrak b-1}}\,
\operatorname{artanh}\!\Bigl(\frac{\sqrt{\mathfrak b}\,\sinh\tau}{\sqrt{\mathfrak b-1}}\Bigr),
\]
and the coefficients \(\mathcal A_{0},\mathcal B_{0},\mathcal A_{1},\mathcal B_{1}, \mathcal A_{2}\) are algebraic
(computable by partial fractions) in \(\tilde\chi,C,\Sigma,D_{a}\) through \(\Lambda_{0,1,2}\).

Let \(W(\tau)\equiv \sqrt{\mathfrak{b}}\,\sinh\tau\). The partial fractions yield
\begin{align} \label{eq:psifun}
\Psi(\tau)=\,
&\frac{\mathcal{A}_{0}}{2\,\Lambda_{2}}\,
\ln\!\bigl(\tilde L(\tau)\bigr)
-\frac{\mathcal{B}_{0}}{2\,\mathfrak{b}}\,
\ln\!\bigl(\mathfrak{b}\cosh^{2}\tau-1\bigr) \\
&\quad+\frac{\mathcal{A}_{1}}{\sqrt{\mathfrak{D}_{\Lambda}}}\,
\arctan\!\Bigl(\frac{2\Lambda_{2}\sinh\tau+\Lambda_{1}}{\sqrt{\mathfrak{D}_{\Lambda}}}\Bigr)
+\frac{\mathcal{B}_{1}}{\sqrt{\mathfrak{b}-1}}\,
\operatorname{artanh}\!\Bigl(\frac{W(\tau)}{\sqrt{\mathfrak{b}-1}}\Bigr)
+\mathcal{A}_{2}\,\tau,
\end{align}
with constants \(\mathcal{A}_{0},\mathcal{B}_{0},\mathcal{A}_{1},\mathcal{B}_{1},\mathcal{A}_{2}\) given explicitly by the (lengthy) algebraic
combinations of \(\tilde\chi,C,\Sigma,D_{\!a}\) that arise in the partial-fraction decomposition of
\(\mathcal{R}/\bigl[(\mathfrak{b}\cosh^{2}\tau-1)\tilde L^{2}\bigr]\).
Since \(\cosh\tau\to 1\) as \(\tau\to 0\) and \(\cosh\tau\sim \frac{1}{2}e^{\tau}\) as \(\tau\to\infty\), all terms have finite limits and the energy is
\begin{equation} \label{eq:enAGD4}
 E_{\mathrm{AGD}}
=\frac{2\,\zeta|\beta|}{\pi\,\kappa^{3}}\,
\Bigl\{\Psi(\infty)-\Psi(0)\Bigr\}.
\end{equation}

\section{Physically relevant solutions}\label{sol}

\begin{figure}
  \centering
   \includegraphics[width=0.5\linewidth]{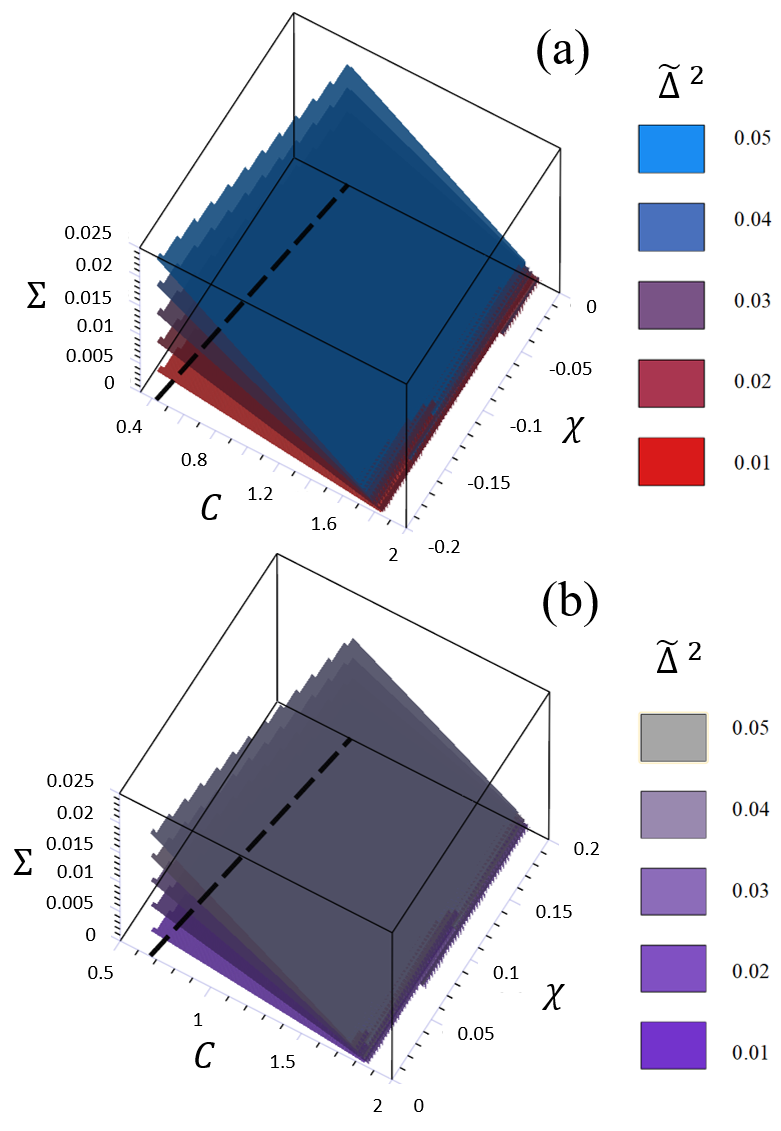}
    \caption{NGD ($\beta>0$, $C>0$) parameter domains for the strongly chirped solution in the three-parameter space $(C,\widetilde\chi,\Sigma)$, shown separately for the two algebraic roots $\widetilde{\Delta}^2_{\pm}$ of Eq.~(15).
The admissible regions are visualized by iso-surfaces $\widetilde{\Delta}^2_{\pm}(C,\Sigma,\widetilde\chi)=\mathrm{const}>0$ subject to criterion~A in Table~1 (positivity of $\widetilde\Delta^2$, $\widetilde{P}_0$, and $\Sigma$).
Panel~(a) shows the plus root $\widetilde{\Delta}^2_{+}$ (displayed for $\widetilde\chi<0$), while panel~(b) shows the minus root $\widetilde{\Delta}^2_{-}$ (displayed for $\widetilde\chi>0$).
Colors correspond to the chosen iso-levels (bar legend). Tildes are omitted from the axis labels for readability. The dashed curve indicates the DSR locus plotted on the reference plane $\Sigma=0$ (see Sec.~4). Note that because $A(\tilde{\chi})=\mathrm{sgn}(\tilde{\chi})\sqrt{Q}$ in Eq.~(\ref{eq:DeltaBranchesCompact}), the one-sided continuations of $\tilde{\Delta}^2_{\pm}$ across $\tilde{\chi}=0$ interchange. This is a reason why the $\tilde{\chi}<0$ and $\tilde{\chi}>0$ sectors are shown separately.
}\label{fig:fig1}
\end{figure}

\begin{figure}
  \centering
   \includegraphics[width=0.6\linewidth]{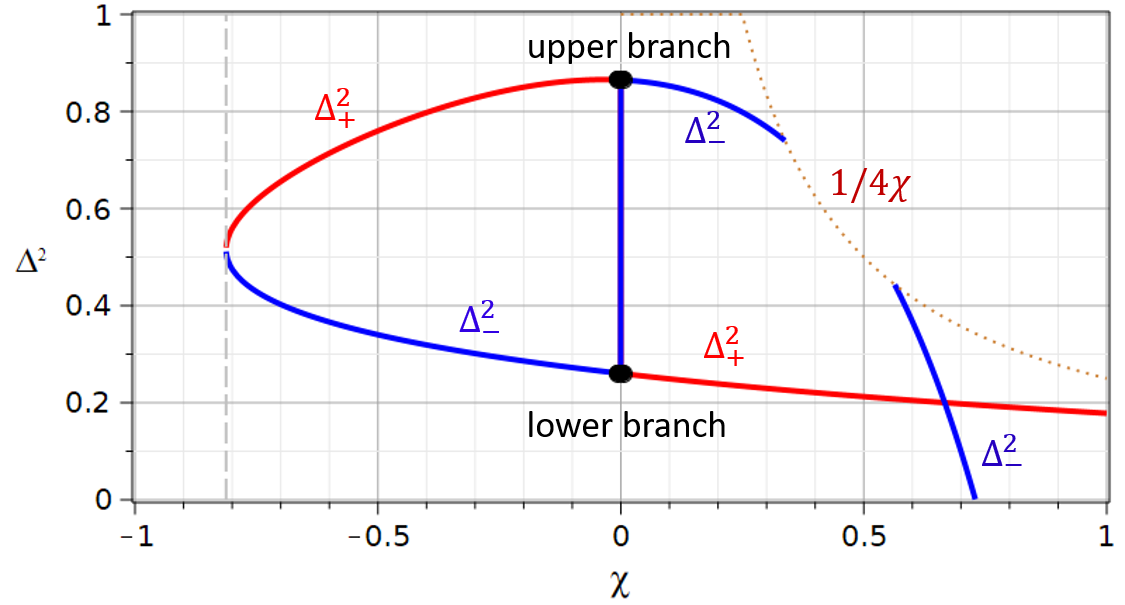}
    \caption{Dependence of $\Delta_{\mp}^2$ on $\chi$ (tildes are removed, the $\Delta_{\mp}^2$-branches are illustrated by the different colors). $C=$0.5, $\Sigma=$0.1.}\label{fig2}
\end{figure}

The physically relevant solutions, i.e., solutions with the positive $\Delta^2$, $P_0$, and $\sigma$, are classified in Table \ref{tbl1} and shown in Figs. \ref{fig:fig1}--\ref{fig:fig3}. The positivity of $\sigma$ is required for vacuum stability of Eq.~(\ref{CGLE}).  The cubic--quintic SAM term in Eq. (\ref{CGLE}) is read as $\kappa \left(1-\zeta P(t)\right)P(t)$, so that $\zeta>0$ prevents the DS collapse with a power growth and provides an energy scaling by the DS broadening and chirp growth. From this point of view, we exclude the B and D sets (Table~\ref{tbl1}) from our further consideration. Simultaneously, the cubic--quintic SPM term in (\ref{CGLE}) is read as $\left(\gamma-\chi P(t)\right) P(t)$. In this case, both signs of the $\chi$-parameter are interesting. For instance, in a laser generating optical pulses by a Kerr-lens mode locking (KLM) mechanism \cite{brabec1992kerr}, a negative $\chi$ could result from a laser mode squeezing due to self-focusing, which is a basic mechanism of KLM. In this case, mode squeezing enhances SPM in a nonlinear medium (SPM self-enhancement). In parallel, a positive $\chi$ results from a SAM saturation ($\zeta>0$) when a laser mode over-squeezing leads to subsequent growth of mode divergence in the nonlinear medium and an effective decrease in SPM (a SPM saturation).

\begin{table}[!t]
\caption{Signs of the dimensionless DS parameters corresponding to the physical solutions, i.e., positive $\Delta^2$, $\sigma$, and $P_0$. The negativeness of $\widetilde{\Delta}^{2}$ means only the negativeness of the corresponding rescaled squared frequency in (\ref{norm}).}\label{tbl1}
\begin{tabular}{llllll}
\toprule
&$\beta$ &  $\zeta$ & $\widetilde{\Delta}^{2}$ & $\widetilde{P}_0$ & $\Sigma$ \\
\midrule
\textbf{A}&>0&>0 & >0 & >0 & >0 \\
\textbf{B}&>0&<0 & <0 & <0 & <0 \\
\textbf{C}&<0&>0 & <0 & >0 & >0 \\
\textbf{D}&<0&<0 & >0 & <0 & <0 \\
\bottomrule
\end{tabular}
\end{table}

\subsection{NGD, $C>0$ (\textbf{A}, Table \ref{tbl1})} \label{cp}
The case of NGD was preliminarily considered in \cite{kalashnikov2009chirped,worksheet,kalashnikov2024dissipative}. Here, we generalize and systematize the obtained results.  Figure~\ref{fig:fig1} visualizes the NGD admissible domains of the strongly chirped DS in the three-parameter space $(C,\Sigma,\widetilde\chi)$.
For clarity, we plot the two algebraic roots $\widetilde{\Delta}^2_{\pm}$ separately: panel~(a) shows iso-surfaces of constant $\widetilde{\Delta}^2_{+}$ in the $\widetilde\chi<0$ sector, while panel~(b) shows iso-surfaces of constant $\widetilde{\Delta}^2_{-}$ in the $\widetilde\chi>0$ sector.
In both cases, the plotted volumes satisfy the criterion~A in Table~1, i.e., $\widetilde{\Delta}^2>0$, $\widetilde{P}_0>0$, and $\Sigma>0$.

Let's introduce the dimensionless quantities
\begin{align}
Q(C,\Sigma,\widetilde{\chi})
\equiv (C-2)^2-16\,\Sigma+16\,C\,\Sigma\,\widetilde{\chi}
=(C-2)^2-16\,\Sigma\,(1-C\widetilde{\chi}),\label{eq:Qdef} \\ 
B\equiv 1-C\widetilde{\chi},\qquad
K\equiv 3-(C+4)\,\widetilde{\chi}, \label{eq:Bdef}
\end{align}
and the polynomial part
\begin{equation}
\mathrm{Poly}(C,\Sigma,\widetilde{\chi})
\equiv -3C+6+\bigl(C^2+2C-8+16\Sigma\bigr)\,\widetilde{\chi}
-16\,C\,\Sigma\,\widetilde{\chi}^{2}.\label{eq:e55}
\end{equation}
With \(A(\widetilde{\chi})=\operatorname{sgn}(\widetilde{\chi})\sqrt{Q}\), the two
branches read
\begin{equation}
\widetilde{\Delta}^{2}_{+}
=\frac{\mathrm{Poly}-K\,A}{8\,B^{2}},
\qquad
\widetilde{\Delta}^{2}_{-}
=\frac{\mathrm{Poly}+K\,A}{8\,B^{2}}.
\label{eq:DeltaBranchesCompact}
\end{equation}

We exclude the pole $B=0$ (i.e., \ $1-C\tilde{\chi}=0$), and require $Q\ge 0$ so that $A$ is real.
Since $8B^2>0$, the sign of $\tilde{\Delta}^2_{\pm}$ is the sign of
$\Phi_{\pm}\equiv \mathrm{Poly}\mp K A$.
The condition $\tilde{P}_{0,\pm}>0$ depends on the sign of $\tilde{\chi}$ (see below).

\paragraph{Case \(\widetilde{\chi}>0\).} Both branches satisfy \(\widetilde{\Delta}^2_{\pm}>0\) and
\(\widetilde{P}^{\pm}_0>0\) if and only if
\begin{equation}
\begin{aligned}
&\text{(A1)}\quad Q(C,\Sigma,\widetilde{\chi})\ \ge\ 0,\\
&\text{(A2)}\quad 1-C\widetilde{\chi}\ \neq\ 0,\\
&\text{(A3)}\quad \mathrm{Poly}(C,\Sigma,\widetilde{\chi})
\ >\ |K|\,\sqrt{Q(C,\Sigma,\widetilde{\chi})},\\
&\text{(A4)}\quad \mathrm{Poly}(C,\Sigma,\widetilde{\chi})
\ +\ |K|\,\sqrt{Q(C,\Sigma,\widetilde{\chi})}
\ \le\ \dfrac{2\,\bigl(1-C\widetilde{\chi}\bigr)^2}{\widetilde{\chi}}\;.
\end{aligned}
\label{eq:chi-positive-box}
\end{equation}

Here (A3) enforces \(\Phi_{\pm}>0\) \emph{simultaneously}. (A4) encodes the
branch wise upper bound \(\widetilde{\Delta}^2_{\pm}\le 1/(4\widetilde{\chi})\),
which is equivalent to \(\widetilde{P}_0^{\pm}>0\) for \(\widetilde{\chi}>0\).

\paragraph{Case \(\widetilde{\chi}<0\).}
Both branches satisfy \(\widetilde{\Delta}^2_{\pm}>0\) and
\(\widetilde{P}^{\pm}_0>0\) if and only if
\begin{equation}
\begin{aligned}
&\text{(B1)}\quad Q(C,\Sigma,\widetilde{\chi})\ \ge\ 0,\\
&\text{(B2)}\quad 1-C\widetilde{\chi}\ \neq\ 0,\\
&\text{(B3)}\quad \mathrm{Poly}(C,\Sigma,\widetilde{\chi})
\ >\ |K|\,\sqrt{Q(C,\Sigma,\widetilde{\chi})}\;.
\end{aligned}
\label{eq:chi-negative-box}
\end{equation}
For \(\widetilde{\chi}<0\), there is no additional upper bound: once
\(\widetilde{\Delta}^2_{\pm}>0\), one has \(\widetilde{P}^{\pm}_0>0\) automatically.

\paragraph{Two–sided limit \(\widetilde{\chi}\to 0^{\pm}\).}
At \(\widetilde{\chi}=0^{\pm}\), the removable singularity is resolved by
\[
\widetilde{\Delta}^2_{s}(0^{\varsigma})=\frac{3}{8}\left[ \left( 2-C \right) -s\varsigma\sqrt{\left( 2-C \right)^2-16\Sigma}\right],
\]
with a reality condition \((2-C)^2-16\Sigma\ge 0\) \cite{podivilov2005heavily} and $s\in \left\{ +1,-1 \right\}$, $\varsigma\in \left\{ +1,-1 \right\}$. \noindent Here $\widetilde{\Delta}_{+}^2$ corresponds $\chi \to 0^{+}$ and $\widetilde{\Delta}_{-}^2$ corresponds $\chi \to 0^{-}$. Thus, in the terms of \cite{kalashnikov2025energy}, $s=-1$, $\varsigma=1$ and $s=1$, $\varsigma=-1$ relate to a \textit{DSR branch} for $\widetilde{\chi} \to 0^{\varsigma}$ and $s=-\sgn{\varsigma}$. This DSR branch is the ``upper branch'' in Fig. \ref{fig2} and extendable to the negative (positive) branch of (\ref{del}) for $\widetilde{\chi}<0$ ($\widetilde{\chi}>0$) in the notations of the present article. Simultaneously, the energy non-scalable branch relates to the ``lower branch'' in Fig. \ref{fig2} and extends to a positive $\widetilde{\Delta}_{+}^2$ (negative $\widetilde{\Delta}_{-}^2$) branch of (\ref{del}) for $\widetilde{\chi}<0$ ($\widetilde{\chi}>0$).

In the vicinity of zero, one may expand (\ref{del}) in the $\widetilde{\chi}$-powers, so that:
\begin{equation}
\widetilde{\Delta}{\mp}^2(\widetilde{\chi})=\sum_{i=0}^{\infty}\widetilde{F}_i^{\mp}\widetilde{\chi}^{i},
\end{equation}
with the first two terms explicitly derived as:
\begin{align}
\widetilde{F}_0^{\mp}=\frac{3}{8}\left[(2 - C)\mp\sqrt{(2 - C)^2-16\Sigma}\right],\\
\widetilde{F}_1^{\mp}=\frac{3(2-C)}{8}\pm\frac{3C\Sigma}{8\sqrt{(2-C)^2-16\Sigma}}-\frac{2C-C^2}{4}\widetilde{F}_0^{\mp}.
\end{align}

The zero-order term corresponds to a strongly chirped DS of the reduced cubic-quintic CGLE in NGD (points in Fig. \ref{fig2}). The general recurrence formula is:
\begin{equation}
\widetilde{F}_i^{\mp} = \frac{a_i}{8}\mp \sum_{k=0}^{i-1}\binom{1/2}{k}\frac{(-1)^k(16C\Sigma)^k}{[(C-2)^2-16\Sigma]^{k+1/2}} b_{i-k} \ -\sum_{j=1}^{\min(i,3)}d_j\widetilde{F}_{i-j}^{\mp},  \quad i \geq 1,
\end{equation}
with the defined coefficients:

\begin{align}
a_i &=[3(2-C),C^2+2C-8+16\Sigma,-16C\Sigma],&\quad a_i &= 0 \quad \text{for } i \geq 4\\
b_i &=[3,\mp(C+4)],& \quad b_i &= 0 \quad \text{for } i \geq 3\\
d_i &=[1,-2(2-C),(2-C)^2], &\quad d_i &= 0  \quad \text{for } i \geq 4.
\end{align}

\begin{figure}
  \centering
   \includegraphics[width=0.8\linewidth]{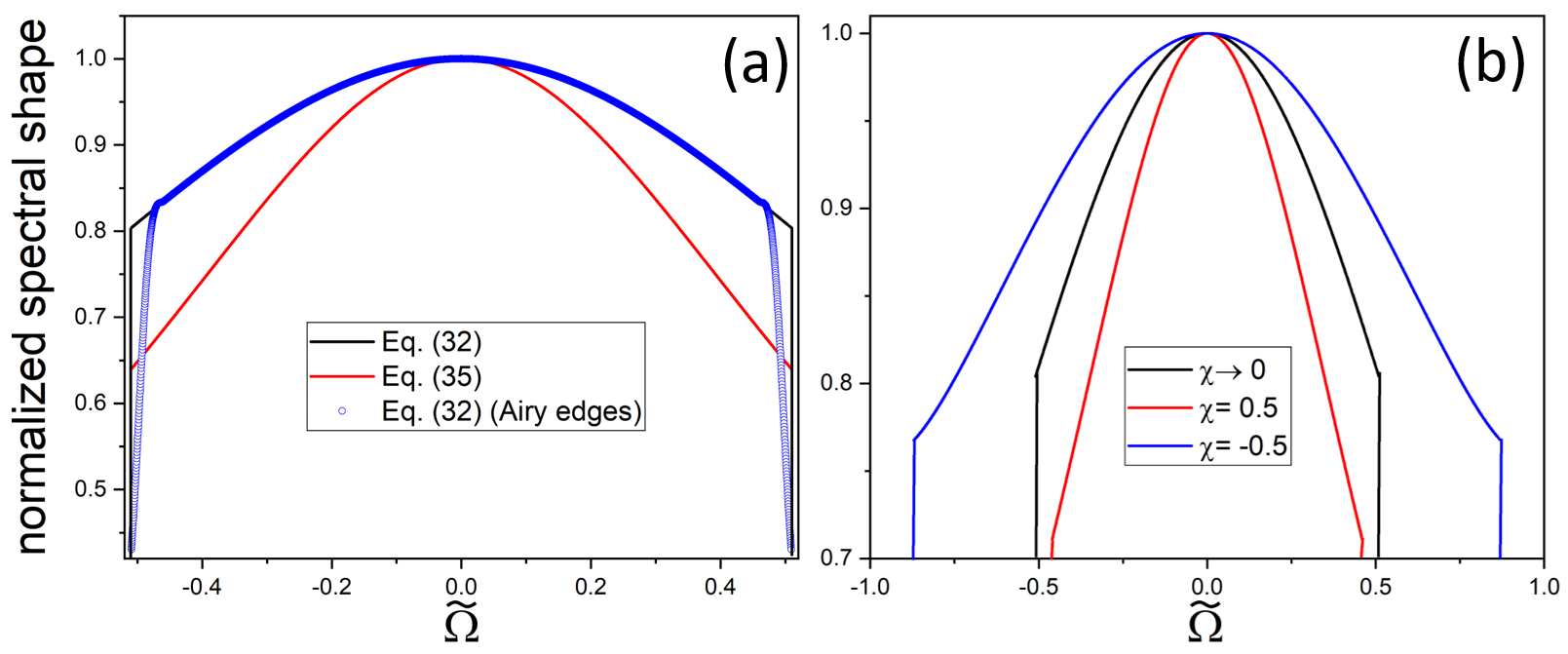}
    \caption{Normalized NGD spectra for the cubic--quintic CGLE at $C=0.5$, $\Sigma=0.1$. (a) Comparison of the quotient-regularized NGD spectrum~(\ref{eq:NGDspa1}) for $\tilde \chi = 10^{-3}$ (blue dashed) with the $\tilde \chi = 0$ limit~(\ref{eq:NGD-Lorentz}) (red solid), which reduces to a truncated Lorentzian $S_{\mathrm{L}} \propto \bigl(\Gamma_{\mathrm{L}}^{2} + \tilde{\Omega}^{2}\bigr)^{-1}$ on $|\tilde{\Omega}| \le \tilde{\Delta}_{0}$. The magenta curve shows the Airy-uniformized version of~(\ref{eq:NGDspa1}), where the Chester--Friedman--Ursell mapping is used to smooth the square-root edges at $|\tilde{\Omega}| \simeq \tilde{\Delta}(\tilde \chi)$ and to remove the spurious node singularities. The three curves almost coincide in the central part of the NGD window, while the Airy uniformization produces only a weak rounding of the cut-off. (b) NGD spectrum~(\ref{eq:NGDspa1}) for three representative values of the chirp-like parameter $\tilde \chi$ (near-NGD, moderate positive and negative $\tilde \chi$), each curve being normalized to its maximum and plotted only inside its own support $|\tilde{\Omega}| \le \tilde{\Delta}(\tilde \chi)$. Increasing $|\tilde \chi|$ changes both the width $\tilde{\Delta}(\tilde \chi)$ and the curvature of the spectrum, but the profile remains close to a truncated Rayleigh--Jeans--type form in the NGD regime \cite{kalashnikov2025energy}.}\label{fig:fig3}
\end{figure}

The NGD spectra in Fig.~\ref{fig:fig3} admit a natural thermodynamic
interpretation. In the limit of $\chi \to 0$, Eq.~(\ref{eq:NGD-Lorentz}) reduces to a Lorentzian
\[
  S_{\mathrm{L}}(\tilde{\Omega}) \propto
  \bigl(\Gamma_{\mathrm{L}}^{2} + \tilde{\Omega}^{2}\bigr)^{-1},
  \qquad |\tilde{\Omega}| \le \tilde{\Delta}_{0}.
\]
That has exactly the structure of a Rayleigh--Jeans (RJ) equilibrium spectrum, i.e.,\ an algebraic decay of the modal occupation number with frequency, truncated by a finite ultraviolet cut-off. Such RJ spectra arise in wave-turbulent thermalization of Hamiltonian nonlinear waves, where the equilibrium mode population is
$n_{k} \propto T/(\varepsilon_{k}-\mu)$ with a high-$k$ cut-off,
see, e.g., \ Ref.~\cite{picozzi2009thermalization}.  In the dissipative cubic-quintic CGLE context, the same truncated-Lorentz profile has been interpreted as the momentum-space RJ distribution of a semi-incoherent DS composed of many weakly interacting ``quasiparticles''~\cite{kalashnikov2025energy}.

Our finite-$\chi$ NGD spectrum~(\ref{eq:NGDspa1}) can therefore be viewed as a controlled deformation of this RJ-like distribution. For small $|\tilde \chi|$, the quotient-regularized global SPA result (blue curve in Fig.~\ref{fig:fig3}(a)) is almost indistinguishable from the truncated Lorentzian in the central region, which means that the semi-incoherent soliton still occupies its spectral modes in an almost thermal (RJ) way. The Airy uniformization only smooths a very thin layer near $|\tilde{\Omega}| \simeq \tilde{\Delta}(\tilde \chi)$, replacing the unphysical square-root singularities by a universal Airy roll-off, while leaving the bulk of the spectrum essentially unchanged. Physically, this corresponds to a slight ``softening'' of the spectral cut-off rather than to a qualitative change of the momentum distribution. Panel \ref{fig:fig3}(b) shows that increasing $|\tilde \chi|$ mainly changes the effective NGD bandwidth $\tilde{\Delta}(\tilde \chi)$ and the curvature of the spectrum, but even for moderately large positive or negative $\tilde \chi$, the profiles remain close to an RJ-type truncated Lorentzian, consistent with the picture of a near-thermal ensemble of modes forming a semi-incoherent DS.

\subsection{AGD, $C<0$ (\textbf{C}, Table \ref{tbl1})} \label{cn}
\begin{figure}[t]
\centering
\includegraphics[width=0.6\linewidth]{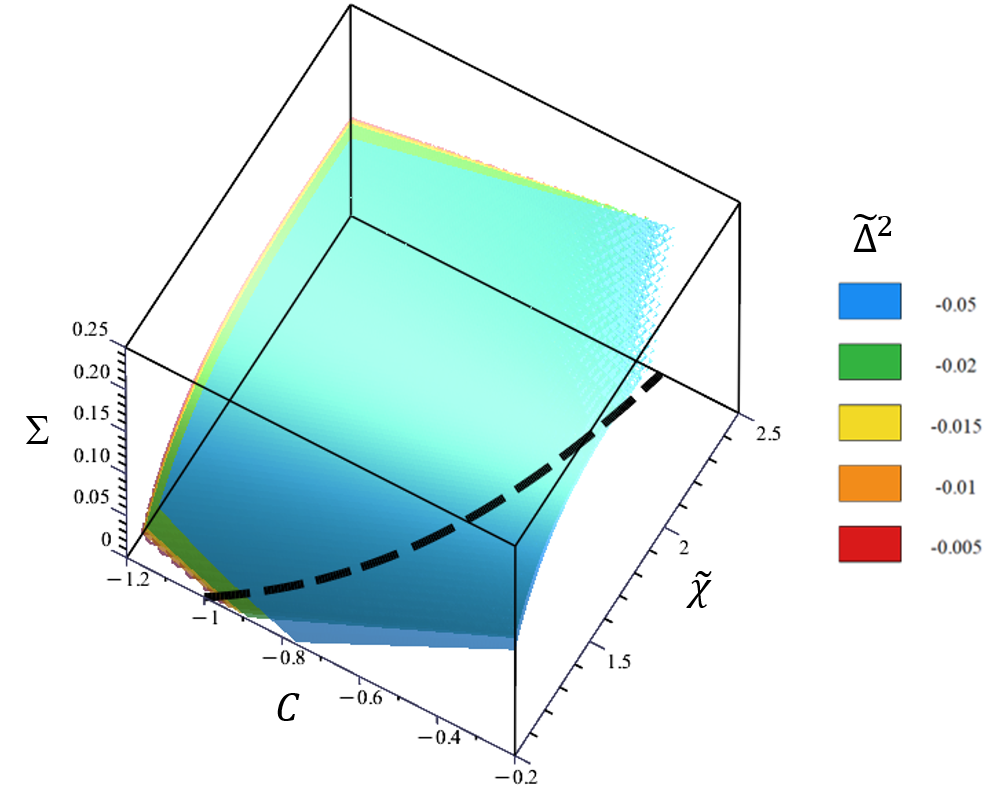}
\caption{AGD parameter domain for the strongly chirped solution (negative branch $\tilde{\Delta}^{\,2}_{-}<0$)
visualized by iso-surfaces of constant $\tilde{\Delta}^{\,2}_{-}(C,\Sigma,\tilde{\chi})=\mathrm{const}<0$
within the AGD adiabatic-existence window \eqref{eq:AGDwindow}.
Colors label the chosen iso-levels (see the bar legend on the right).
The dashed curve in the $(C,\tilde{\chi})$ base plane marks the AGD--DSR resonance line $1+C\tilde{\chi}=0$
(i.e.\ $C=-1/\tilde{\chi}$; cf.\ Sec.~5), indicating where the AGD master-diagram energy divergence can occur, provided that the locus lies within the AGD adiabatic-existence window.}
\label{fig4}
\end{figure}

The strongly chirped DS in the AGD exists only for a negative branch $\widetilde{\Delta}^{2}_-$ and positive $\widetilde{\chi}$, which means a SPM saturation with a power (see Fig. \ref{fig4}). Formally, one must use the other quadratic root for $\widetilde{P}_0$, that is the solution without the $\chi \to 0$ limit (compare with (\ref{pp})):
\begin{equation}
    \widetilde{P}_0= \frac{1+\sqrt{1-4\widetilde{\chi}\widetilde{\Delta}_{-}^2}}{2\widetilde{\chi}}, \label{eq:P0AGD}
\end{equation}
\noindent existing only for $\widetilde{\chi}>0$ to satisfy the criterion \textbf{C} in Table \ref{tbl1}. 

Figure~\ref{fig4} summarizes the admissible AGD domain for the strongly chirped DS on the minus branch,
$\tilde{\Delta}^{\,2}_{-}<0$ \footnote{Let's remind that the negativeness of $\tilde{\Delta}^{\,2}$ means only the negativeness of the scaling coefficient in the frequency normalization in (\ref{norm}).}.
Rather than plotting only the boundary of the admissible set, we visualize the interior of the admissible volume by several
iso-surfaces $\tilde{\Delta}^{\,2}_{-}(C,\Sigma,\tilde{\chi})=\mathrm{const}<0$.
For any fixed pair $(C,\tilde{\chi})$ in the AGD sector ($C<0$, $\tilde{\chi}>0$), the strongly chirped adiabatic solution
persists only while $\tilde{\Delta}^{\,2}_{-}$ remains negative, i.e., for $\Sigma$ satisfying the AGD adiabatic-existence constraint $\Sigma<\Sigma_{\max}(C,\tilde\chi)$.
The colored sheets, therefore, provide a direct geometric picture of how the negative-$\tilde{\Delta}^{\,2}_{-}$ branch
fills the three-parameter space and how the admissible volume evolves with $C$ and $\tilde{\chi}$.

The dashed curve in the $(C,\tilde{\chi})$ base plane marks the AGD--DSR resonance locus $1+C\tilde{\chi}=0$
(i.e., $C=-1/\tilde{\chi}$; see Sec.~5).
Its position relative to the projected AGD window anticipates an important qualitative distinction from NGD: the DSR-type energy divergence can occur only when this resonance line intersects the AGD existence domain, which requires sufficiently large $\tilde{\chi}$.

Let us consider the limitations on the parameter space analytically. With the definitions
\begin{equation} \label{eq:BdefAGD}
    B = 1 - C\,\widetilde{\chi}, \qquad K = 3 - (C+4)\,\widetilde{\chi},
\end{equation}
\begin{equation}
Q=(C-2)^{2}-16\,\Sigma\,(1-C\tilde\chi),
\label{eq:QdefAGD}
\end{equation}
\begin{equation}
\mathrm{Poly}(C,\Sigma,\chi)=-3C+6+(C^{2}+2C-8+16\Sigma)\tilde\chi-16C\Sigma\tilde\chi^{2},
\label{eq:PolyDefAGD}
\end{equation}
\noindent the minus branch reads
\begin{equation}
\tilde\Delta_{-}^{2}=\frac{\mathrm{Poly}+K\sqrt{Q}}{8B^{2}}\ (<0\ \text{in AGD}).
\label{eq:DeltaMinusAGD}
\end{equation}
In the AGD setting, we use (\ref{eq:P0AGD}),
so that for \(\widetilde{\chi}>0\), the condition \(\widetilde{P}_0>0\) is automatically met once \(\widetilde{\Delta}^{2}_{-}<0\).
Since \(8B^2>0\) for \(C<0,\ \widetilde{\chi}>0\), the sign of \(\widetilde{\Delta}^{2}_{-}\) is the sign of
\(\Phi_{-}\equiv  \mathrm{Poly} + K\sqrt{Q}\).
The boundary \(\Phi_{-}=0\) yields the closed-form threshold
\begin{equation}
\Sigma_{\rm th}(C,\tilde\chi)=\frac{3\big(\tilde\chi(C+4)-3\big)}{16\tilde\chi^{2}},
\label{eq:Sth}
\end{equation}

Thus, for a fixed \((C,\widetilde{\chi})\) with \(C<0\) and \(\widetilde{\chi}>0\),
\begin{equation}
\label{eq:phys-region}
\widetilde{\Delta}^{2}_{-}<0
\quad \Longleftrightarrow \quad
0<\Sigma<\min\!\big\{\Sigma_{\mathrm{th}}(C,\widetilde{\chi}),\ \Sigma_{Q}(C,\widetilde{\chi})\big\},
\end{equation}
together with the reality constraint
\begin{equation}
\label{eq:Sigma-Q}
\Sigma_{Q}(C,\widetilde{\chi}) \;=\; \frac{(C-2)^2}{16\,(1 - C\widetilde{\chi})} \quad \text{(from $Q\ge 0$)}.
\end{equation}

\begin{itemize}
\item \textit{Onset in \(\widetilde{\chi}\).} A negative-\(\widetilde{\Delta}^{2}_{-}\) region exists iff \(\Sigma_{\mathrm{th}}>0\),
i.e.
\begin{equation}
\widetilde{\chi} > \frac{3}{C+4} \qquad \text{(requires $C>-4$)}.
\end{equation}

\item \textit{Upper bound on $\Sigma$.} Maximizing \(\Sigma_{\mathrm{th}}(C,\widetilde{\chi})\) over \(\widetilde{\chi}>0\) gives
\begin{equation}
\label{eq:Sigma-max}
\Sigma_{\max}(C) \;=\; \max_{\widetilde{\chi}>0}\Sigma_{\mathrm{th}}(C,\widetilde{\chi})
\;=\; \frac{(C+4)^2}{64}\;
\quad\text{attained at}\quad
\widetilde{\chi} \;=\; \frac{6}{C+4}\,.
\end{equation}

\item \textit{Inverted threshold (fixed \(\Sigma\)).} Solving \(\Sigma=\Sigma_{\mathrm{th}}(C,\tilde\chi)\) for \(\widetilde{\chi}\) yields
\begin{equation}
\label{eq:chi-min}
\widetilde{\chi} \;\ge\; \widetilde{\chi}_{\min}(C,\Sigma)
\;=\; \frac{3\Big((C+4)-\sqrt{(C+4)^2-64\,\Sigma}\Big)}{32\,\Sigma}
\end{equation}
\noindent which is real only for \(\Sigma \le (C+4)^2/64\), and must be \noindent combined with \eqref{eq:Sigma-Q}:
\end{itemize}

\begin{equation}
C<0,\ \tilde\chi>0,\ B>0,\ Q\ge 0,\ \tilde\Delta_{-}^{2}<0,\
0<\Sigma<\min\{\Sigma_{\rm th}(C,\tilde\chi),\,\Sigma_{Q}(C,\tilde\chi)\}\,
\label{eq:AGDwindow}
\end{equation}

\begin{figure}
    \centering
    \begin{subfigure}{0.45\textwidth}
    \includegraphics[width=1\linewidth]{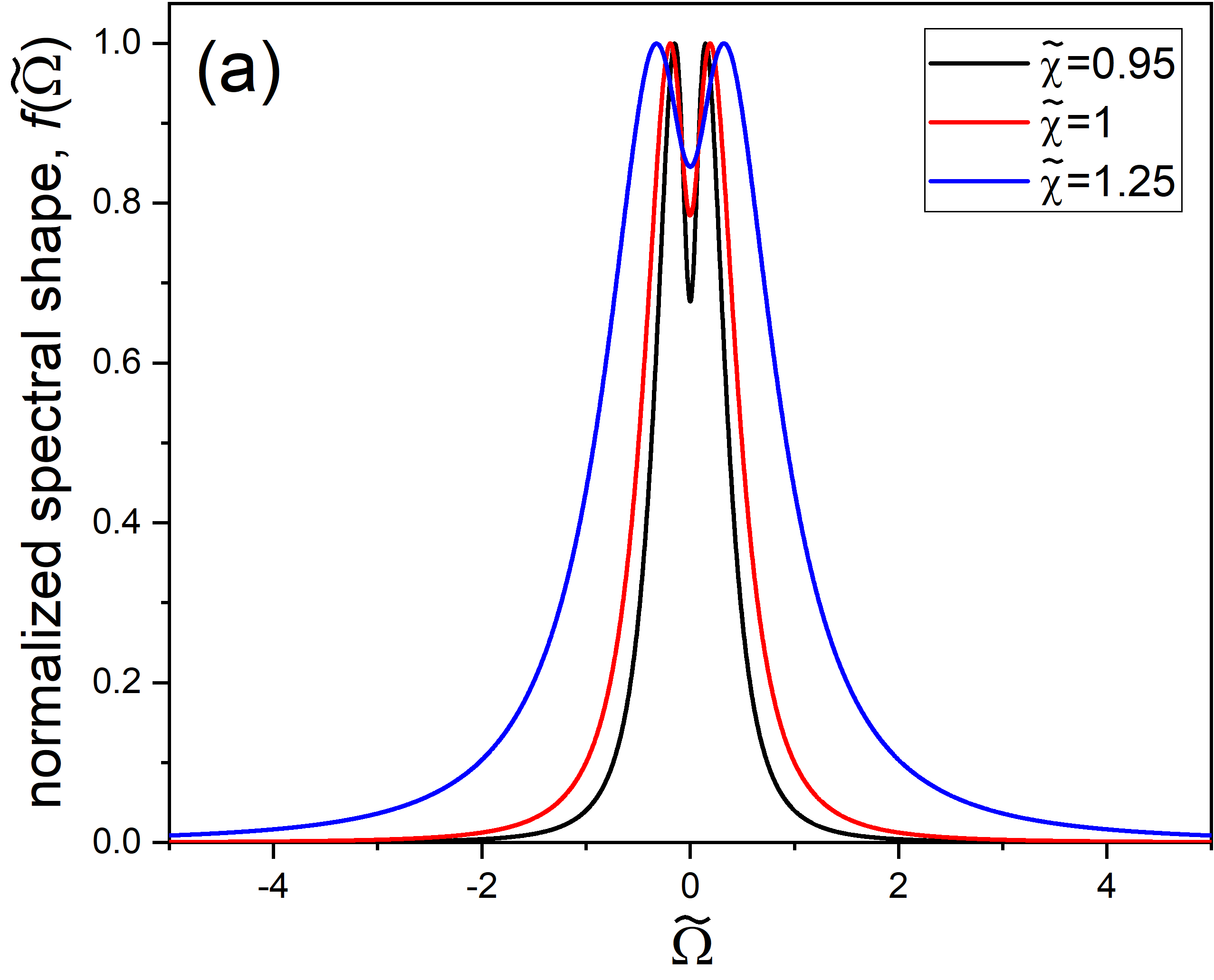}
    \end{subfigure}
        \begin{subfigure}{0.45\textwidth}
    \includegraphics[width=1\linewidth]{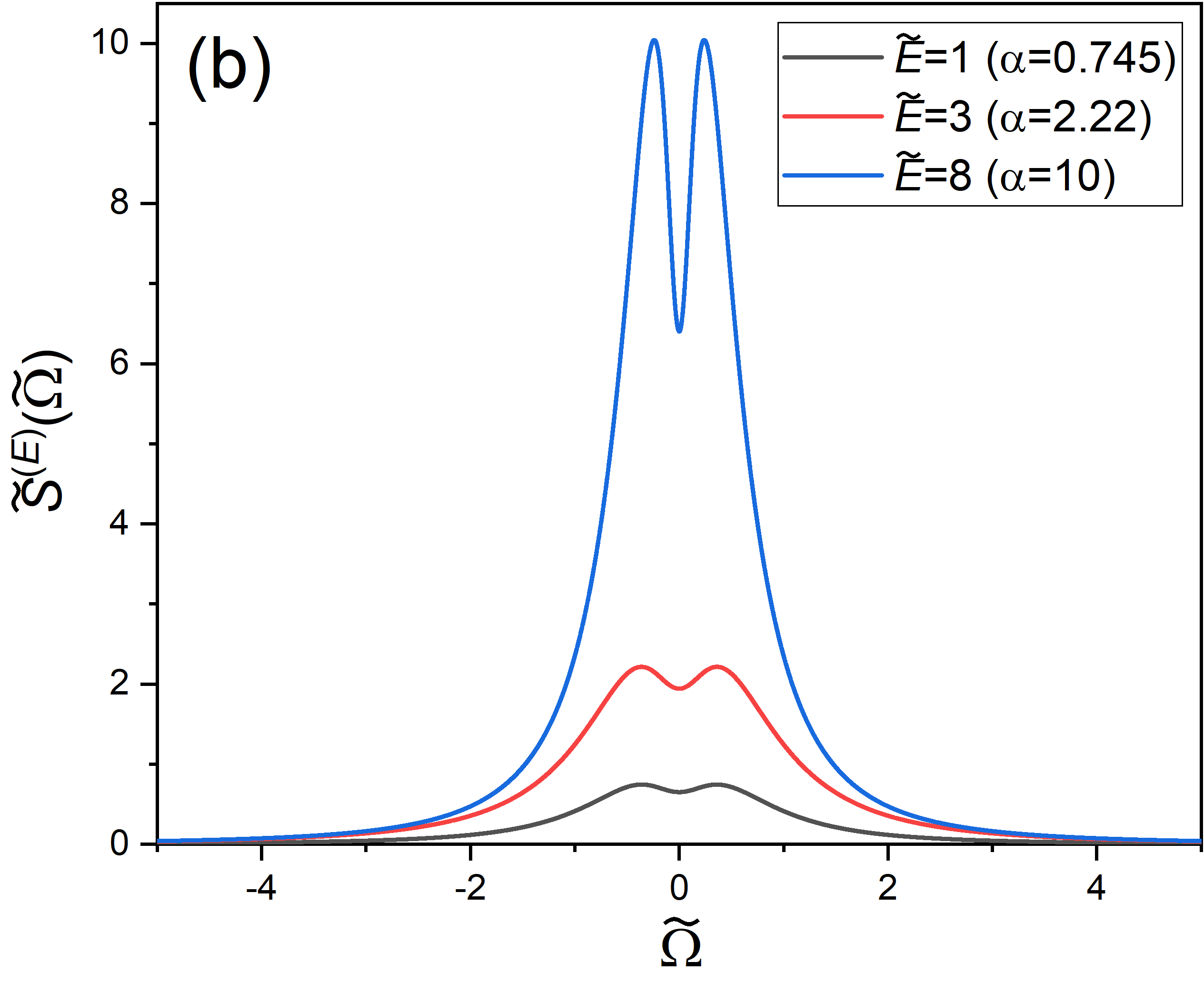}
    \end{subfigure}
    \caption{(a) Stationary–phase AGD spectra for the cubic–quintic CGLE at
$C=-0.5$ and $\Sigma=0.05$.  Normalized spectral envelopes
$\hat{S}_{\rm AGD}(\tilde\Omega)$ computed from Eq.~(\ref{eq:AGDspa1}) for three values of the normalized quintic SPM parameter $\tilde\chi=0.95$ (black), 1 (red), and 1.25 (blue). The spectra are scaled to unit peak for shape comparison; the central depression becomes shallower as $\tilde\chi$ increases. (b) Energy scaling near the AGD--DSR line.
Unit-peak envelopes $f(\tilde\Omega)=\hat S_{\rm AGD}(\tilde\Omega)/\max_{\tilde\Omega}\hat S_{\rm AGD}$
(from Eq.~(44)) are rescaled as $\hat S^{(E)}(\tilde\Omega)=\xi(\tilde E)\,f(\tilde\Omega)$ with
$\int_{0}^{\tilde\Omega_{\rm cap}}\hat S^{(E)}(\tilde\Omega)\,d\tilde\Omega=\tilde E$ ($\tilde\Omega_{\rm cap}=6$).
Parameters: $\tilde\chi=1.5$, $C=-0.668667$ ($C=-1/\tilde\chi+\delta C$, $\delta C=-0.002$); $\tilde E=1,3,8$.}
    \label{fig:fig5}
\end{figure}

The representative stationary–phase spectra in the AGD regime are shown in Fig.~\ref{fig:fig5} for $C=-0.5$, $\Sigma=0.05$, and three values of the normalized quintic SPM parameter $\tilde\chi>0$.  The spectra $\hat{S}_{\mathrm{AGD}}(\tilde\Omega)$ are obtained from the AGD SPA
expression~(\ref{eq:AGDspa1}) and normalized to unit peak for shape comparison. All curves exhibit a smooth central spectral depression and slowly decaying wings. Moreover, the depth of the depression decreases with increasing $\tilde\chi$, in agreement with the discussion above.  For large
$|\tilde\Omega|$, the spectra follow the algebraic tail $\hat{S}_{\mathrm{AGD}}(\tilde\Omega)\propto|\tilde\Omega|^{-3}$ given
by Eq.~(\ref{eq:AGDspa2}), confirming that the pulses are strongly chirped in this
parameter regime. 
To relate the unit-peak envelope shapes in Fig.~\ref{fig:fig5}(a) to the energy scalability in the AGD master diagram (Fig.~\ref{fig:DSRmaster:AGD}),
we introduce an energy-scaled spectrum $\hat S^{(E)}(\tilde\Omega)=\xi(\tilde E)\,f(\tilde\Omega)$,
where $f(\tilde\Omega)$ is the unit-peak envelope, and $\xi(\tilde E)$ is fixed by
$\int_{0}^{\tilde\Omega_{\rm cap}}\hat S^{(E)}(\tilde\Omega)\,d\tilde\Omega=\tilde E$
(here $\tilde\Omega_{\rm cap}$ is a finite plotting window used only for visualization in Fig.~\ref{fig:fig5}(b)).
A representative near-resonant example (fixed $\tilde\chi$ and $C\simeq-1/\tilde\chi$) is shown in Fig.~\ref{fig:fig5}(b),
illustrating that $f(\tilde\Omega)$ depends only weakly on $\tilde E$, while the energy variation is primarily carried by
the scalar factor $\xi(\tilde E)$.

In the strongly chirped regime, the parameters $(C,\Sigma,\tilde{\chi})$ satisfy the AGD adiabatic-existence window  (cf. Eqs.~(\ref{eq:QdefAGD})–(\ref{eq:Sigma-Q})):

\begin{equation}
C<0,\quad \tilde\chi>0,\quad B>0,\quad Q\ge0,\quad
\tilde\Delta_-^{\,2}<0,\quad
0<\Sigma<\min\{\Sigma_{\mathrm{th}}(C,\tilde\chi),
               \Sigma_Q(C,\tilde\chi)\},
\label{eq:AGDwindow}
\end{equation}

For fixed $(C,\Sigma)$, this implies a lower bound $\tilde\chi\ge\tilde\chi_{\min}(C,\Sigma)$ for the existence of a strongly–chirped AGD DS.  At $C=-0.5$ and $\Sigma=0.05$, one finds $\tilde\chi_{\min}\simeq0.92$, so only the
cases with $\tilde\chi\gtrsim0.95$ in Fig.~\ref{fig:fig5} belong to this adiabatic AGD branch.  Here, the instantaneous frequency is large and monotonic, the stationary–phase approximation leading to Eq.~(\ref{eq:SPA_power_explicit2}) (or its
normalized form~(\ref{eq:AGDspa1})) is valid, and the spectra show the characteristic
central depression with $|\tilde\Omega|^{-3}$ tails.

\subsubsection{Coherent multi-horn spectra in AGD}
\label{sssec:coh-multihorn-agd}

The SPA yields the spectral field as a coherent two-saddle sum
(Eq.~(\ref{eq:SPA_amplitude})) with amplitudes and phases defined by Eqs.~(\ref{eq:SPA_phase},\ref{eq:27}).
The envelope spectra used throughout this work correspond to \emph{fringe averaging} of the
$\pm$ interference term in the power spectrum, as stated explicitly in Eq.~(\ref{eq:SPA_power_general}) and summarized by the
envelope form (\ref{eq:env-times-fringe}). In the AGD regime ($\beta<0$, $C<0$, $\tilde\chi>0$), the resulting explicit
envelope is given by Eq.~(\ref{eq:AGDspa1}).

In fully coherent numerical simulations, or in experiments without sufficient ensemble/time averaging,
the $\pm$ interference is not removed. Retaining it gives the coherent (non-averaged) spectral power
in the standard two-saddle form
\begin{equation}
\hat S^{(\mathrm{coh})}_{\mathrm{AGD}}(\tilde\Omega)
=
\hat S^{(\mathrm{env})}_{\mathrm{AGD}}(\tilde\Omega)
\Bigl[1+V(\tilde\Omega)\cos\Delta\Phi(\tilde\Omega)\Bigr],
\qquad
V(\tilde\Omega)=\frac{2|A_+(\tilde\Omega)A_-(\tilde\Omega)|}{|A_+(\tilde\Omega)|^2+|A_-(\tilde\Omega)|^2}\in[0,1],
\label{eq:ScohAGD}
\end{equation}
where $\hat S^{(\mathrm{env})}_{\mathrm{AGD}}(\tilde\Omega)$ is the fringe-averaged SPA envelope
(Eqs.~(\ref{eq:SPA_power_general},~\ref{eq:env-times-fringe},~\ref{eq:AGDspa1})), $A_\pm$ are the SPA amplitudes from Eq.~(\ref{eq:27}),
and $\Delta\Phi(\tilde\Omega)\equiv\Phi_+(\tilde\Omega)-\Phi_-(\tilde\Omega)$ is the relative phase of
the two stationary contributions computed from Eq.~(\ref{eq:SPA_phase}) using the AGD chirp law $d\tilde\Omega/d\tilde t$ in Eq.~(\ref{eq:omega_sweep_explicit}).

For a symmetric single-hump pulse, one typically has $|A_+|\simeq|A_-|$ in the central band, hence
$V(\tilde\Omega)\simeq 1$ there. While in the far wings one saddle dominates and $V(\tilde\Omega)\to 0$,
so that $\hat S^{(\mathrm{coh})}_{\mathrm{AGD}}\to \hat S^{(\mathrm{env})}_{\mathrm{AGD}}$ asymptotically.
The interference factor in Eq.~\eqref{eq:ScohAGD} can therefore (i) preserve a two-horn profile when
$\cos\Delta\Phi$ is non-constructive near $\tilde\Omega\simeq 0$, or (ii) fill the smooth central depression of the
envelope and produce a stable three-horn pattern when the interference becomes sufficiently constructive in the
central band. Because both $V(\tilde\Omega)$ and $\Delta\Phi(\tilde\Omega)$ vary smoothly with the parameters,
the transition between two- and three-horn spectra is continuous.

The horn positions are governed by the constructive-interference condition
$\Delta\Phi(\tilde\Omega)=2\pi m$ (local maxima for $V$ are not too small), while the local minima satisfy
$\Delta\Phi(\tilde\Omega)=(2m+1)\pi$.
This coherent-fringe mechanism is independent of the NGD node regularization discussed above. 
\footnote{Eq.~(\ref{eq:GammaL}) concerns the NGD node-window width in the $\chi\to 0$ limit and should not be interpreted as an
AGD envelope expression.}

When the parameters are changed such that, at fixed $(C,\tilde\chi)$,
the threshold $\Sigma_{\mathrm{th}}(C,\tilde\chi)$ is crossed, the minus root changes sign, $\tilde\Delta_-^{\,2}>0$, and the inequalities in \eqref{eq:AGDwindow} are no longer satisfied.  The strongly chirped AGD solution then ceases to exist, but the cubic-quintic CGLE still supports localized DSs.  These states form a different, \emph{weakly chirped, soliton–like} AGD branch lying outside the adiabatic domain (\ref{eq:ScohAGD}), where the adiabatic assumptions no longer hold. Their spectra deviate qualitatively from Eq.~(\ref{eq:AGDspa1}): the central dip is
filled in and the wings develop an oscillatory, almost sinc–like
structure, as seen in the full numerical simulations.

In this weakly chirped regime, the pulse intensity profile is well approximated by the Pereira--Stenflo-type profile of Ref.~\cite{pereira1977nonlinear}, which also underlies the exact cubic-quintic CGLE solitary-wave ansatz used in Ref.~\cite{renninger2008dissipative},
\[
  a(t)\;\propto\;\frac{1}{A+\cosh(t/T)}\,\exp[i\phi(t)],
\]
which reduces to a $\sech$ pulse only in the limit $A\to0$, while for
finite $A>0$ it has a flat–topped (table–top) shape with a sinc–like
spectrum.  In the terminology of the present work, this Pereira--Stenflo-type profile should be regarded as a \emph{morphological}
approximation for the weakly chirped cubic-quintic CGLE solitons that occur once the AGD adiabatic window~(\ref{eq:AGDwindow}) is violated: the central depression is filled in, and the wings develop oscillations, in contrast to the strongly chirped AGD branch
described by the SPA envelope spectra in Fig.~\ref{fig:fig5}.
A closely related \emph{exact} chirped ansatz of the Pereira--Stenflo/Hocking--Stewartson type was employed in Ref.~\cite{akhmediev1995novel} (codimension-one constraint manifold), and is invoked here only as a
useful reference for the weakly-chirped/table-top pulse morphology, not as an AGD asymptotic formula.

\section{DSR in NGD for finite quintic SPM}
\label{sec:DSR-NGD}

In Refs.~\cite{kalashnikov2024dissipative,kalashnikov2025energy}, the DSR in the NGD regime was defined in terms of the ``master diagram'' as the 
existence of an energy-scalable branch of strongly chirped DSs whose 
energy diverges in the vacuum-instability limit $\Sigma \to 0^+$ at fixed 
control parameter $C$. In the reduced CGLE with vanishing quintic SPM ($\tilde\chi = 0$), this corresponds to the upper branch of the NGD solution on the $(C,\Sigma)$-plane, where the Lorentzian cutoff width $\Delta$ saturates, the effective ``chemical potential'' $\Xi$ tends to zero, and the energy $E$ exhibits an infinite asymptotics in a finite interval $C \in [C_-,C_+]$ on the master diagram.%

In the present notation, the $\tilde\chi \to 0^\varsigma$ limit of the NGD
cutoff (Eq.~(\ref{del})) is described by the expansions
\begin{equation}
  \Delta^2_{\mp}(\tilde\chi)
  = \sum_{i=0}^{\infty} \tilde F^{(\mp)}_i \, \tilde\chi^i,
  \qquad
  \tilde F^{(\mp)}_0
  = \frac{3}{8}\Big[(2-C) \mp \sqrt{(2-C)^2 - 16\Sigma}\Big],
  \label{eq:delta_expansion_DSR_comment}
\end{equation}
with the first $\tilde\chi$–correction given by Eq.~(\ref{eq:chi-positive-box}).%
\footnote{As discussed below, the DSR branch at $\tilde\chi \to 0^\varsigma$ corresponds
to $s = -\mathrm{sgn}\,\varsigma$ in the notation of Eq.~(\ref{eq:e55}), and matches the 
energy-scalable branch in the master diagram of Ref.~\cite{kalashnikov2024dissipative}.}
For $\tilde\chi = 0$, the NGD spectrum reduces to the node-free truncated Lorentzian
\begin{equation}
  \hat S^{(0)}_{\mathrm{NGD}}(\tilde\Omega)
  =
  \mathcal{A}_L\,\frac{\Gamma_L^2}{\Gamma_L^2 + \tilde\Omega^2}\,
  \Theta(\Delta^2 - \tilde\Omega^2),
  \label{eq:lorentz_DSR_comment}
\end{equation}
with cutoff $\Delta$ and Lorentzian width $\Gamma_L$ given by Eqs.~(\ref{eq:energy1})--(\ref{eq:NGDspa1}),
and the corresponding NGD energy is
\begin{equation}
  E^{(0)}_{\mathrm{NGD}}
  = \frac{\mathcal{A}_L}{\pi}\,\arctan\!\left(\frac{\Delta}{\Gamma_L}\right)
  \label{eq:energy_chi0_DSR_comment}
\end{equation}
(cf.~Eq.~(\ref{eq:NGDen1})). On the DSR branch, the ratio $\Delta/\Gamma_L$ and the effective chemical potential $\Xi$ behaves such that the energy displays 
an infinite asymptotics $E^{(0)}_{\mathrm{NGD}} \to \infty$ as 
$\Sigma \to 0^+$ for $C$ inside the DSR interval of the master diagram, 
while $\Delta$ saturates and the spectrum approaches a ``finger-like'' 
truncated Rayleigh–Jeans profile~\cite{kalashnikov2024dissipative,kalashnikov2025energy}.

For finite quintic SPM ($\tilde\chi \neq 0$), the node-regularized NGD energy is 
given by Eqs.~(\ref{eq:tildeP_explicit2})--(\ref{eq:s0ngd}), and the elementary closed form obtained in Eqs.~(\ref{eq:NGDen2})--(\ref{eq:NGD-Lorentz}):
\begin{equation}
  E_{\mathrm{NGD}}(\tilde\chi)
  = \frac{2\zeta|\beta|}{\pi\kappa^{3}}\,|\theta|
  \begin{cases}
    \big[\Phi_{\arctan}(y)\big]_{y_L}^{y_U}, & \mathcal{D}>0,\\[1ex]
    \big[\Phi_{\artanh}(y)\big]_{y_L}^{y_U}, & \mathcal{D}<0,
  \end{cases}
  \label{eq:ENGD_finite_chi_DSR_comment}
\end{equation}
where $\mathcal{D} = 4\mathfrak{A}\mathfrak{G} - \mathfrak{B}^2$ is the discriminant of the 
quadratic denominator in the $y$–variable, with coefficients $(\mathfrak{A},\mathfrak{B},\mathfrak{G})$ 
defined in Eq.~(\ref{eq:NGDen2}). The finite–$\tilde\chi$ NGD spectrum remains a controlled 
deformation of the truncated Rayleigh–Jeans profile~(\ref{eq:energy1}), with a weak Airy 
rounding of the cutoff and a $\tilde\chi$–dependent bandwidth $\Delta(\tilde\chi)$ 
(see Fig.~\ref{fig:fig3} and the discussion around Eq.~(61)).

In this setting, we adopt the same operational definition of DSR: DSR corresponds to the 
existence, for fixed $(C,\tilde\chi)$, of an energy–scalable NGD branch such that
\begin{equation}
  \exists\,C^*(\tilde\chi):\quad
  \lim_{\Sigma\to 0^+} E_{\mathrm{NGD}}(C^*,\Sigma,\tilde\chi) = \infty,
  \label{eq:DSR_def_finite_chi}
\end{equation}
while the spectral support $\Delta(C^*,\Sigma,\tilde\chi)$ remains finite and the effective chemical potential tends to zero, producing a finger-like spectrum.
Using Eq.~(\ref{del}) for the NGD cutoff and the small–$\tilde\chi$ expansion 
(\ref{eq:delta_expansion_DSR_comment}), one finds that the DSR branch at $\tilde\chi\neq 0$ 
is obtained by a regular continuation of the $\tilde\chi=0$ DSR branch into the 
$(C,\Sigma,\tilde\chi)$-space: the infinite-energy asymptotics in 
Eq.~(\ref{eq:DSR_def_finite_chi}) persists, and the DSR interval in $C$ is deformed
only perturbatively,
\begin{equation}
  C_{\mathrm{DSR},\pm}(\tilde\chi)
  = C_{\mathrm{DSR},\pm}(0)
    + \delta C_{\pm}\,\tilde\chi
    + \mathcal{O}(\tilde\chi^{2}),
  \label{eq:C_DSR_shift}
\end{equation}

\noindent where $C_{\mathrm{DSR},\pm}(0)$ are the NGD DSR boundaries of the reduced
($\tilde\chi = 0$) master diagram~\cite{kalashnikov2024dissipative,kalashnikov2025energy}. The coefficients
$\delta C_{\pm}$ can be obtained explicitly from the implicit DSR condition by
combining the expansion of $\Delta^2(C,\Sigma,\tilde\chi)$ (Eqs.~(\ref{eq:e55})--(\ref{eq:chi-positive-box}))
with the existence constraints (A1)–(A4)/(B1)–(B3) in Eqs.~(\ref{eq:Qdef})--(\ref{eq:Bdef}), but the resulting expressions are algebraically cumbersome and do not yield a simple 
closed form. In practice, the $\tilde\chi$–dependence of the DSR window in $C$ 
is most conveniently evaluated numerically from Eq.~(\ref{del}) together with the 
positivity constraints for $\Delta^2$ and $\tilde P_0$.

\begin{figure*}[t]
\centering
\begin{subfigure}[t]{0.45\textwidth}
  \centering
  \includegraphics[width=\linewidth]{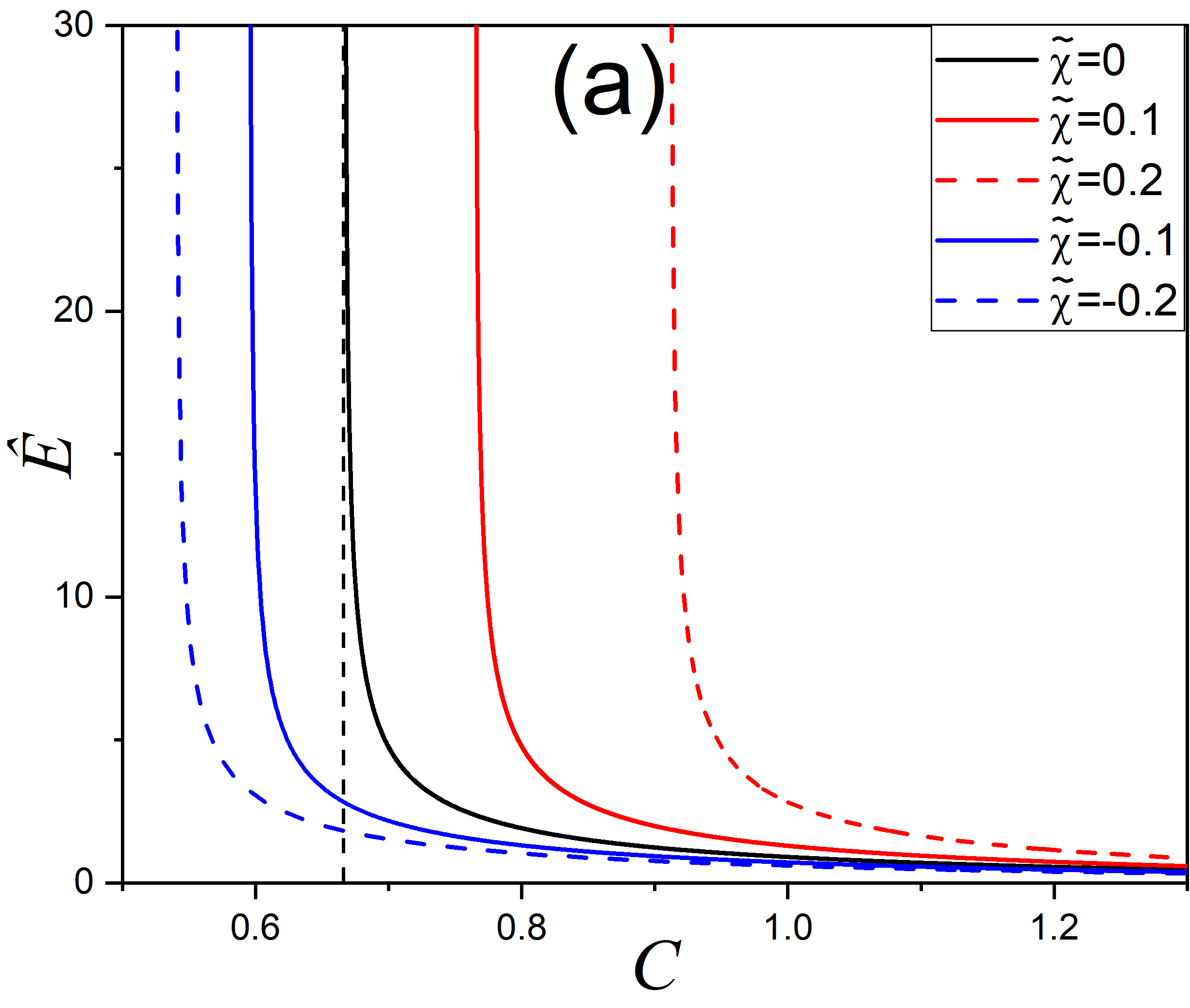}
  \caption{NGD master diagram (\(C>0\)).}
  \label{fig:DSRmaster:NGD}
\end{subfigure}\hfill
\begin{subfigure}[t]{0.45\textwidth}
  \centering
  \includegraphics[width=\linewidth]{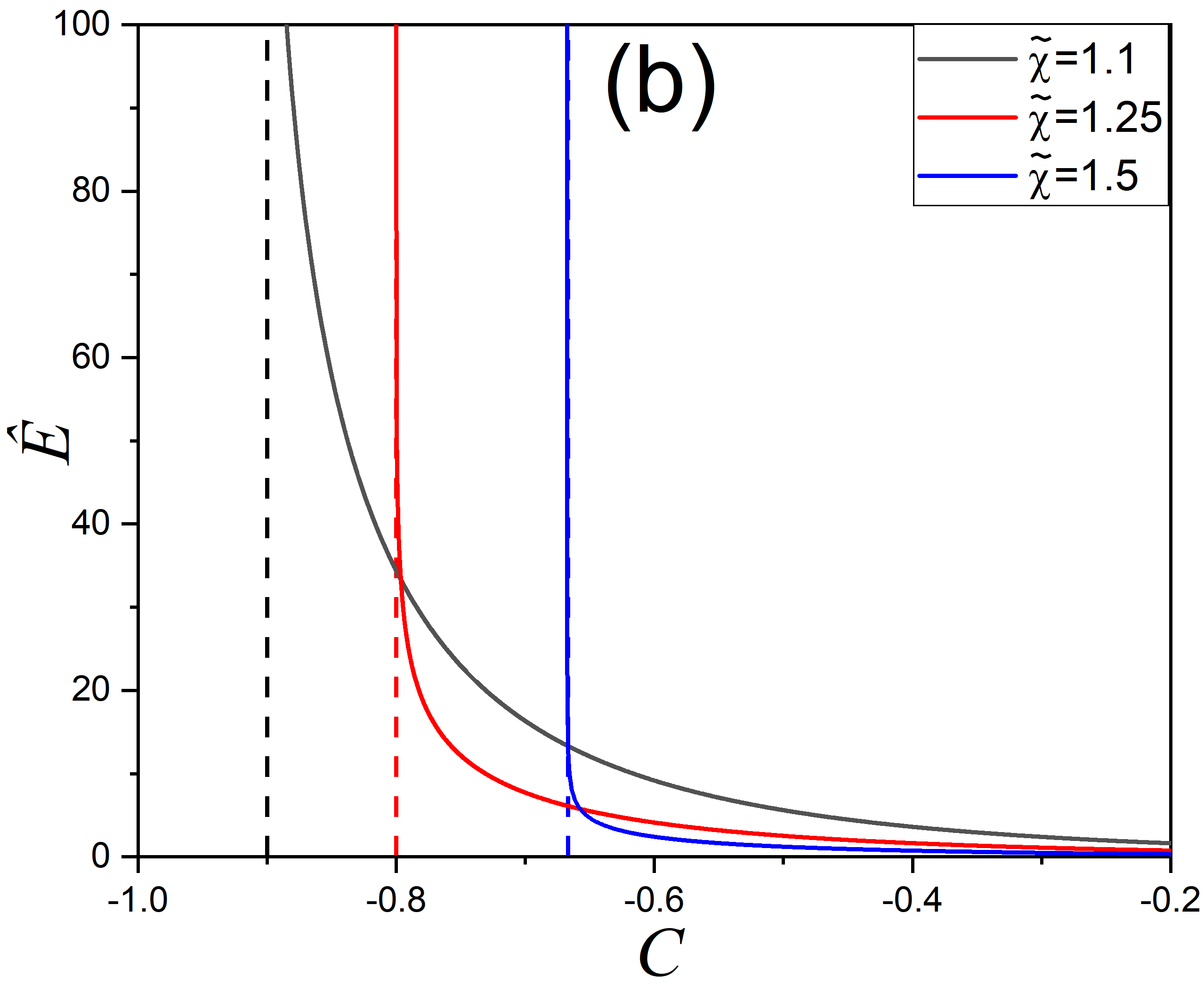}
  \caption{AGD master diagram (\(C<0\)).}
  \label{fig:DSRmaster:AGD}
\end{subfigure}
\caption{DSR master diagrams:
(a) NGD: the normalized energy \(\tilde E(C)\) for \(\Sigma\to 0^+\) at several values of the normalized quintic SPM parameter \(\tilde\chi\). The \(\tilde E\) behavior demonstrates DSR where an energy diverges; the vertical black dashed line indicates the DSR limit \(C_{\mathrm{DSR}}(0)=2/3\) for $\tilde\chi=0$. The DS exists on the left side of the corresponding curves (i.e., for smaller $C$) where \(\Sigma > 0\).
(b) AGD: \(\tilde E\) develops a vertical asymptote at the AGD--DSR locus \(C_{\mathrm{DSR}}^{(\mathrm{AGD})}(\tilde\chi)=-1/\tilde\chi\) (vertical dashed lines), provided this point lies inside the AGD adiabatic-existence window.}
\label{fig:DSRmaster}
\end{figure*}

To visualize how a finite quintic SPM modifies the NGD DSR branch, we evaluate the normalized NGD energy $\tilde{E}$ obtained from Eq.~(\ref{eq:ENGD_finite_chi_DSR_comment}) in the limit of $\Sigma \to 0^{+}$ and plot its dependence on the control parameter $C$ for several representative values of the normalized quintic SPM parameter $\tilde{\chi}$ (see Fig.~\ref{fig:DSRmaster:NGD}).
For each fixed $\tilde\chi$, the curve in Fig.~\ref{fig:DSRmaster:NGD} marks the vacuum-stability boundary ($\Sigma\to0^{+}$). The physically admissible strongly chirped DSs occupy the region at smaller $C$ (where one can take $\Sigma>0$), whereas larger $C$ would correspond to $\Sigma<0$ and background growth. The vertical dashed line indicates the cubic-SPM limit ($\tilde\chi=0$) reference position, while finite $\tilde\chi$ shifts the divergence along the $C$-axis. For $\tilde\chi=0$, the curve reproduces the reduced CGLE result: $\tilde E$ diverges as $C\to C_{\mathrm{DSR}}=2/3$, while for $C<C_{\mathrm{DSR}}$ the energy remains finite and the strongly chirped DS is vacuum-stable, which is provided by $\Sigma>0$. Thus, DS is stable on the left side of the corresponding curve in Fig. ~\ref{fig:DSRmaster:NGD}.

Finite quintic SPM preserves this
energy-scalable behavior, but slightly shifts the position of the resonance:
positive $\tilde{\chi}$ (saturable SPM) displaces the divergence towards larger $C$, whereas
negative $\tilde{\chi}$ (self-enhancing SPM) shifts it to smaller $C$. Away from the resonance the normalized energy $\tilde{E}(C,\tilde{\chi})$ decays monotonically with $C$. The curves for different $\tilde{\chi}$ remain close to each other, indicating that the quintic SPM mainly deforms the boundaries of the DSR window rather than the global shape of the scalable branch. This numerical continuation confirms that the DSR interval $C_{\mathrm{DSR},\pm}(\tilde{\chi})$ obtained from Eq.~(\ref{del}) is a smooth perturbation of the cubic DSR interval, consistent with the expansion (\ref{eq:C_DSR_shift}), while the explicit analytical expressions for the shifts $\delta C_{\pm}$ remain algebraically unwieldy and are best handled numerically.

\section{DSR in AGD for finite quintic SPM}
\label{sec:DSR-AGD}

In Sec.~\ref{sec:DSR-NGD} we formulated DSR in NGD via the master diagram \(\tilde E(C)\) at fixed \(\Sigma\) as the existence of a branch whose energy diverges as \(\Sigma\to0^+\) for some \(C\). At the same time, the solution remains within the adiabatic DS family.
Here we apply the same operational definition to the AGD (\(C<0\)) but with an important structural difference: in AGD, the SPA spectrum does not terminate at a finite cut-off frequency. Instead, it tends to an arbitrarily large $|\tilde\Omega|$ and decays algebraically.

Following Eq.~(\ref{eq:DSR_def_finite_chi}), we say that DSR occurs in AGD if there exists a value \(C_{\mathrm{DSR}}^{(\mathrm{AGD})}(\tilde\chi)\) such that the (adiabatic) AGD energy diverges in the vacuum-stability limit,
\begin{equation}
\exists\, C_{\mathrm{DSR}}^{(\mathrm{AGD})}(\tilde\chi):\qquad
\lim_{\Sigma\to0^+}\tilde E_{\mathrm{AGD}}\!\left(C_{\mathrm{DSR}}^{(\mathrm{AGD})},\Sigma,\tilde\chi\right)=\infty,
\label{eq:AGD-DSR-def}
\end{equation}
while the solution remains on the AGD adiabatic branch (in particular, \(\tilde\Delta_-^2<0\)).

The key observation comes from the large-\(|\tilde\Omega|\) asymptotics of the AGD SPA spectrum: away from the special parameter values it decays as \(|\tilde\Omega|^{-3}\) with a prefactor controlled by the ``chirp--control'' combination \((1+C\tilde\chi)\) (cf. Eq.~(\ref{eq:AGDspa2})).
Consequently, the AGD master-diagram energy becomes singular when \((1+C\tilde\chi)\to0\): in this limit, the leading term used to obtain Eq.~(\ref{eq:AGDspa2}) is no longer uniform, the tail amplitude blows up, and, at the resonance point, the next-order balance yields a nonintegrable contribution. That identifies the AGD DSR locus as the chirp--control line
\begin{equation}
1+C\tilde\chi=0
\qquad\Longrightarrow\qquad
C_{\mathrm{DSR}}^{(\mathrm{AGD})}(\tilde\chi)=-\frac{1}{\tilde\chi}.
\label{eq:AGD-DSR-locus}
\end{equation}

However, Eq.~\eqref{eq:AGD-DSR-locus} produces an actual DSR divergence only if the resonance point lies \emph{inside} the AGD adiabatic-existence window of the adiabatic solution.
In the AGD regime, this window is defined by Eq.~(\ref{eq:AGDwindow}), which includes \(\tilde\Delta_-^2<0\).
In the limit \(\Sigma\to0^+\) the boundary of the AGD branch is given by \(\tilde\Delta_-^2(C,0,\tilde\chi)=0\), yielding the lower-\(C\) boundary
\begin{equation}
C_{\min}(\tilde\chi)=\frac{3}{\tilde\chi}-4,
\label{eq:Cmin-chi}
\end{equation}
so that, for \(\Sigma\to0^+\), the adiabatic AGD branch exists for \(C\in\bigl(C_{\min}(\tilde\chi),0\bigr)\).
Requiring the resonance point \(C_{\mathrm{DSR}}^{(\mathrm{AGD})}=-1/\tilde\chi\) to belong to this interval gives
\begin{equation}
-\frac{1}{\tilde\chi}>\frac{3}{\tilde\chi}-4
\qquad\Longleftrightarrow\qquad
\tilde\chi>1.
\label{eq:AGD-DSR-threshold}
\end{equation}

Thus, unlike NGD (where DSR already exists in the cubic limit), AGD-DSR requires a sufficiently strong quintic SPM contribution:
\begin{equation}
\tilde\chi_{\min}=1,
\qquad
C_{\mathrm{DSR}}^{(\mathrm{AGD})}(\tilde\chi)=-\tilde\chi^{-1},
\qquad (\Sigma\to0^+).
\label{eq:AGD-DSR-summary}
\end{equation}

For numerical master diagrams, \(\Sigma\) is small but finite, which shifts the onset slightly upward.
Expanding the AGD boundary condition \(\tilde\Delta_-^2\!\left(C=-1/\tilde\chi,\Sigma,\tilde\chi\right)=0\) near \(\tilde\chi=1\) yields
\begin{equation}
\tilde\chi_{\min}(\Sigma)\simeq 1+\frac{4}{3}\Sigma,
\qquad \Sigma\ll1,
\label{eq:AGD-DSR-finiteSigma}
\end{equation}
that displays an appearance of the divergence at finite \(\Sigma\).

Figure~\ref{fig:DSRmaster:AGD} illustrates these conclusions in a form which is directly comparable to the NGD master diagram of Fig.~\ref{fig:DSRmaster:NGD}. The solid curves again show the normalized master-diagram energy $\tilde E$ as a function of the chirp-control parameter $C$ at a fixed positive $\Sigma$, now on the AGD branch ($C<0$). However, the meaning of the curves differs from that of panel (a).
In NGD (Sec.~\ref{sec:DSR-NGD}), the master diagram is naturally read as a stability boundary because the physically admissible chirped DS requires a stable zero background, i.e., $\Sigma>0$. In AGD, by contrast, we plot $\tilde E(C)$ along the strongly chirped branch at a fixed small positive $\Sigma$, so the zero background is already stable and does not mean that the curves define the DS stability boundaries. Instead, when $C$ is moved too far (toward $0$ or too negative, depending on $\tilde\chi$), one of the basic admissibility requirements of the adiabatic branch fails (the algebraic ``chirped-soliton'' solution ceases to be real/physical). Therefore, the endpoints (dashed curves $C_{\mathrm{DSR}}^{(\mathrm{AGD})}=-1/\tilde\chi$) of the AGD curves in Fig.~\ref{fig:DSRmaster:AGD} should be read as the limits of the adiabatic existence window of the strongly chirped AGD branch in $C$, not as a change of the background-stability condition.

As shown by Chang \textit{et al.}~\cite{chang2009dissipative}\footnote{We use the different sign notations in the cubic-quintic CGLE (\ref{CGLE}): $D>0$ (Chang et. al.) corresponds to our $\beta<0$ (AGD), $\nu<0$ corresponds to $\tilde\chi>0$, $\epsilon > 0$ to $\kappa>0$, and $\mu>0$ to $\kappa \zeta <0$.}, who computed resonance curves and existence regions for high-energy DSs of the cubic-quintic CGLE, the resonance curve can be continued into the AGD domain for a saturable SPM (positive $\tilde \chi$). In their simulations, the pulse energy grows rapidly near a DSR in the AGD regime, and the corresponding spectra develop pronounced two-sided maxima as the resonance is approached, which is qualitatively consistent with the structured AGD spectral core predicted by the present stationary-phase theory. At the same time, Chang \textit{et al.} emphasize that the steepness (``effectiveness'') of the resonance is strongly controlled by the quintic nonlinear term, rather than being a generic consequence of AGD alone (like \cite{duan2011experimental}). The master-diagram viewpoint adopted here clarifies this point geometrically. In AGD, the DSR-type divergence is not automatic, but occurs only when the resonance locus intersects the physically admissible strongly-chirped AGD existence window (hence, there is a threshold on $\tilde\chi$). At the same time, once this intersection exists, the approach to the resonance can indeed produce a very sharp increase in energy.

The weakly-chirped AGD branch discussed above (outside the adiabatic window (\ref{eq:AGDwindow})) is closely related, at the level of
pulse morphology, to the exact stationary CGLE solitary waves considered in \cite{akhmediev1995novel} for AGD and in \cite{renninger2008dissipative} for NGD regimes.  There, the field is represented by an explicit chirped ansatz of the Pereira--Stenflo/Hocking--Stewartson type,
which yields a broad class of flat-top pulses and also pulses with steep spectral edges and a dip in the center.
However, that construction is a codimension one, i.e., inserting the ansatz into the CGLE produces algebraic constraints
among the governing coefficients, and the existence of the analytic pulse requires a parameter relation (a constraint manifold)
rather than an open existence domain.

By contrast, the present adiabatic theory for strongly chirped DSs provides a \emph{codimension--zero}
description within its validity domain: the solitary pulse exists throughout the AGD adiabatic--existence window (\ref{eq:AGDwindow}),
which can be visualized as a three--dimensional region in $(C,\Sigma,\tilde\chi)$ (Fig.~\ref{fig4}), and yields universal master--diagram
predictions for the energy scalability (Fig.~\ref{fig:DSRmaster}).  In particular, the AGD--DSR is identified by the
chirp--control condition $1+C\tilde\chi=0$. An actual divergence of the master diagram $\tilde E(C)$ (i.e., DSR) occurs only when this
resonance locus lies inside the window (\ref{eq:AGDwindow}), which requires sufficiently large quintic SPM (Sec.~\ref{fig:fig5} and Fig.~\ref{fig:DSRmaster:AGD}).

Consequently, the exact cubic-quintic CGLE pulses of Refs.~\cite{akhmediev1995novel,chang2009dissipative} and the adiabatic strongly--chirped AGD branches described here should be
viewed as complementary limits within the broader CGLE phenomenology of AGD DS: outside (\ref{eq:AGDwindow}), one naturally encounters weakly--chirped,
table--top/oscillatory spectra, while inside (\ref{eq:AGDwindow}) the stationary--phase spectra exhibit the characteristic smooth depression and
$|\tilde\Omega|^{-3}$ tails (Fig.~\ref{fig:fig5}), and the DSR condition becomes a transparent geometric statement in parameter space
(Figs.~\ref{fig4} and \ref{fig:DSRmaster}).

\section{Autocorrelation and two-scale microstate picture in AGD}
\label{sssec:ACF-AGD}

\begin{figure}[t]
  \centering
  \includegraphics[width=\linewidth]{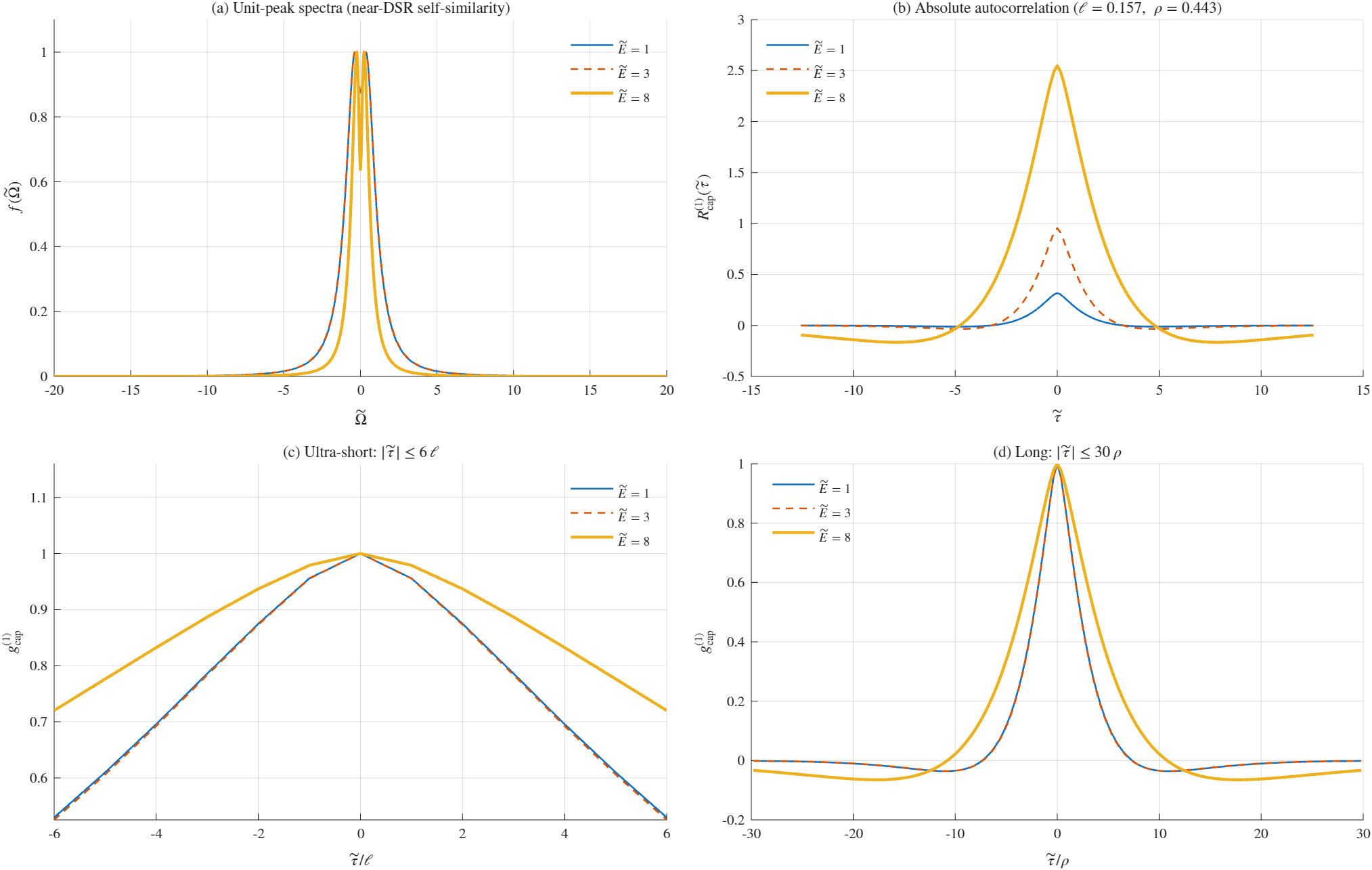}
  \caption{Spectral self-similarity and two correlation times in the strongly--chirped AGD regime near the AGD--DSR line.
The field autocorrelation is computed from the \emph{windowed} AGD SPA spectrum
$\hat S^{(\mathrm{cap})}_{\mathrm{AGD}}(\tilde\Omega)=\hat S_{\mathrm{AGD}}(\tilde\Omega)\,H(\tilde\Omega_{\mathrm{cap}}-|\tilde\Omega|)$,
using the cosine transform
$R^{(1)}_{\mathrm{cap}}(\tilde\tau)=\pi^{-1}\int_{0}^{\infty}\hat S^{(\mathrm{cap})}_{\mathrm{AGD}}(\tilde\Omega)\cos(\tilde\Omega\tilde\tau)\,d\tilde\Omega$
and the normalized coherence
$g^{(1)}_{\mathrm{cap}}(\tilde\tau)=R^{(1)}_{\mathrm{cap}}(\tilde\tau)/R^{(1)}_{\mathrm{cap}}(0)$.
\textbf{(a)} Unit-peak windowed spectra
$f(\tilde\Omega)=\hat S^{(\mathrm{cap})}_{\mathrm{AGD}}(\tilde\Omega)/\max_{\tilde\Omega}\hat S^{(\mathrm{cap})}_{\mathrm{AGD}}(\tilde\Omega)$
illustrating near-DSR spectral self-similarity (collapse of the spectral core) for different $\tilde E$.
\textbf{(b)} Absolute $R^{(1)}_{\mathrm{cap}}(\tilde\tau)$ showing that the \emph{magnitude} scales with energy
via the near-DSR prefactor $\xi(\tilde E)$ in Eq.~(96).
\textbf{(c)} The same normalized coherence in the ultra-short variable $\tilde\tau/\ell$,
where the short correlation scale $\ell\sim\pi/\tilde\Omega_{\mathrm{cap}}$ is set by the effective bandwidth
(sinc-graining kernel).
\textbf{(d)} The coherence in the long variable $\tilde\tau/\rho$,
where $\rho\sim\tilde\Omega_{\mathrm{core}}^{-1}$ is controlled by the spectral core (Eqs.~(105)--(106)).
Parameters correspond to the near-resonant example of Fig.~5(b): $\tilde\chi=1.5$,
$C\simeq-1/\tilde\chi+\delta C$ with $\delta C=-0.002$; curves shown for $\tilde E=1,3,8$.
(For the plotted example $\tilde\Omega_{\mathrm{cap}}=20$, hence $\ell=\pi/\tilde\Omega_{\mathrm{cap}}=0.157$; the core criterion yields $\rho\simeq0.443$.)}
  \label{fig:agd_autocorr_two_scales}
\end{figure}

The near-DSR self-similarity of the unit-peak windowed AGD envelope spectra (Fig.~\ref{fig:agd_autocorr_two_scales}a) motivates considering the temporal-coherence properties of strongly chirped DSs in the AGD regime. In the NGD regime, the autocorrelation of a strongly chirped DS was shown to possess two distinct
correlation scales, enabling an interpretation of the DS as a composite of interacting ``quasiparticles''
(``microstates'') confined by a common potential \cite{akhmediev2000multi,picozzi2007towards,picozzi2009thermalization,picozzi2014optical}\footnote{In some sense, such a structure can be treated as a ``composite soliton'' \cite{soto2001interrelation,soto2002composite}.}, and thereby allowing a statistical-thermodynamic description
of chirped DSs near DSR \cite{kalashnikov2025energy}.

\paragraph{Field autocorrelation from the AGD SPA spectrum.}
In the AGD adiabatic regime ($\beta<0$, $C<0$, $\tilde\chi>0$, $\tilde\Delta_-^2<0$), the SPA provides the
explicit envelope spectrum $\hat S_{\rm AGD}(\tilde\Omega)$ (\ref{eq:AGDspa1}), which has no strict cut-off and decays as
$\hat S_{\rm AGD}\sim|\tilde\Omega|^{-3}$ (\ref{eq:AGDspa2}), so that the normalized energy is finite and given by Eq.~(\ref{eq:enAGD}).
Since $\hat S_{\rm AGD}(\tilde\Omega)$ is an even function, the (first-order) field autocorrelation can be written as
\begin{equation}
R^{(1)}_{\rm AGD}(\tilde\tau)
\equiv
\frac{1}{\pi}\int_{0}^{\infty}\hat S_{\rm AGD}(\tilde\Omega)\cos(\tilde\Omega \tilde\tau)\,d\tilde\Omega,
\qquad
g^{(1)}_{\rm AGD}(\tilde\tau)=\frac{R^{(1)}_{\rm AGD}(\tilde\tau)}{R^{(1)}_{\rm AGD}(0)}.
\label{eq:R1_AGD_def}
\end{equation}
The corresponding intensity autocorrelation can be obtained from $g^{(1)}$ in the usual manner if the field
fluctuations are close to Gaussian (Siegert-type relation), which is the natural setting for statistical interpretations.

Fig.~\ref{fig:agd_autocorr_two_scales} makes explicit the logical chain from spectra to coherence.
Panel~(a) demonstrates the near-DSR \emph{shape} invariance by removing the trivial energy factor through unit-peak normalization,
$f(\tilde\Omega)\equiv\hat S^{(\mathrm{cap})}_{\mathrm{AGD}}(\tilde\Omega)/\max \hat S^{(\mathrm{cap})}_{\mathrm{AGD}}$.
The corresponding energy scaling is then restored by
$\hat S^{(\tilde E)}_{\mathrm{AGD}}(\tilde\Omega)\approx \xi(\tilde E)\,f(\tilde\Omega)$,
with $\xi(\tilde E)$ fixed by the windowed area constraint (cf. Fig.~\ref{fig:fig5}(b)).
Substitution into the cosine transform immediately explains the separation visualized in Fig.~\ref{fig:agd_autocorr_two_scales}(b--d) and summarized by Eq.~(\ref{eq:scaling_corr}) (see below):
the absolute autocorrelation scales as $R^{(1)}_{\mathrm{cap}}(\tilde\tau;\tilde E)\propto\xi(\tilde E)$,
whereas the normalized coherence $g^{(1)}_{\mathrm{cap}}(\tilde\tau)$ is governed primarily by the common spectral core and reveals two distinct
correlation scales, $\ell$ and $\rho$.

\paragraph{Origin of two correlation scales in AGD.}\footnote{For a detailed derivation of the windowed spectrum, convolution form, and the short/long correlation scales, see Appendix~C.}
A key difference between NGD and AGD is the absence of a hard spectral cut-off in AGD: the envelope is unbounded.
However, a spectral dissipation is essential for DS formation in the NGD regime \cite{Chong:08}, and, generally, a momentum cutoff is crucial for the thermalization of incoherent nonlinear waves \cite{picozzi2009thermalization}. In our case of AGD, we could assume
$\tilde\Omega_{\rm cap}$ used in Fig.~\ref{fig:fig5}(b) to be $\approx 2\pi/\sqrt{\alpha}$ (see Eq. (\ref{CGLE})\footnote{The squared inverse bandwidth of the spectral filter $\alpha$ must be normalized in agreement with (\ref{norm}).}) so that the spectrum of DS is windowed: 
$\hat S_{\rm AGD}^{(\rm cap)}(\tilde\Omega)=\hat S_{\rm AGD}(\tilde\Omega)\,H(\tilde\Omega_{\rm cap}-|\tilde\Omega|)$.
By the convolution theorem\footnote{For a more detailed derivation and asymptotic estimates (including the wing contribution), see Appendix~C.},
\begin{equation}
R^{(1)}_{\rm AGD,cap}(\tilde\tau)
=
R^{(1)}_{\rm AGD}(\tilde\tau)\ast
\left[
\frac{\tilde\Omega_{\rm cap}}{\pi}\,\mathrm{sinc}\!\left(\tilde\Omega_{\rm cap}\tilde\tau\right)
\right],
\label{eq:R1_convolution}
\end{equation}
one may introduce a \emph{short} correlation scale
\begin{equation}
\ell \;\sim\; \frac{\pi}{\tilde\Omega_{\rm cap}}
\label{eq:ell_micro_AGD}
\end{equation}
(the ``graining'' scale in the sense of \cite{kalashnikov2025energy}). This short-scale behavior has a direct analogue in a weakly dissipative BEC: in driven/dissipative condensates, first-order correlation functions are commonly evaluated with an explicit ultraviolet (high-frequency) cutoff to control short-time/short-distance divergences, which primarily affects temporal correlations. In particular, de~Leeuw \textit{et al.} emphasize that a finite lifetime has a strong effect on temporal $g^{(1)}$ and note that a cutoff regularization is routinely introduced in nonequilibrium-condensate correlation calculations \cite{de2014phase}.
Accordingly, in our AGD setting, $\tilde\Omega_{\rm cap}$ should be read as an effective spectral dissipation window, and $\ell$ as the associated resolution-limited correlation time.

Independently, the \emph{long} correlation scale is set by the width of the central spectral core of
$\hat S_{\rm AGD}$, which can be characterized, for example, by an effective core bandwidth $\tilde\Omega_{\rm core}$
defined via a cumulative-energy fraction,
$\int_{0}^{\tilde\Omega_{\rm core}}\hat S_{\rm AGD}d\tilde\Omega=(1-\varepsilon)\int_{0}^{\infty}\hat S_{\rm AGD}d\tilde\Omega$
with a fixed $\varepsilon\ll1$.
This yields
\begin{equation}
\varrho \;\sim\; \tilde\Omega_{\rm core}^{-1},
\label{eq:rho_potential_AGD}
\end{equation}
which controls the broad pedestal of $|g^{(1)}_{\rm AGD}(\tilde\tau)|$ and can be interpreted as the width of the
collective confining ``potential'' for microscopic degrees of freedom.

Thus, even without a strict spectral cut-off, AGD admits the same two-scale hierarchy $(\ell \ll \varrho)$ emphasized in the NGD thermodynamic picture~\cite{kalashnikov2025energy}
\footnote{We stress that this BEC connection is made at the level of coherence structure under a finite bandwidth (cut-off) and the resulting scale separation, rather than implying identical microscopic kinetics.}. Here, the \emph{short} time $\ell\sim \pi/\tilde\Omega_{\rm cap}$ is imposed by the finite spectral dissipation/filtering (``UV cutt-off''), while the \emph{long} time $\varrho\sim \tilde\Omega_{\rm core}^{-1}$ is controlled by the intrinsic low-frequency (``IR'') spectral core.

The emergence of distinct coherence scales is not unique to mode-locked lasers or to the cubic--quintic CGLE. It is a generic feature of driven open condensates whose dynamics admit an effective GP/CGLE-type description, including exciton--polariton and photon condensates (with intrinsic drive and lifetime) and atomic condensates with weak loss or reservoir coupling. For quantum fluids of light, the standard mean-field framework is a driven--dissipative GP equation with explicit loss and drive~\cite{CarusottoCiuti2013}, while modern field-theoretic formulations emphasize that stationary states generically violate detailed balance and that coherence is jointly controlled by drive and dissipation~\cite{Sieberer2025RMP}. For cold atoms with inelastic processes, weak two-body loss can be incorporated at the mean-field level by a complex interaction constant (complex scattering length), leading to a dissipative GP equation of the form $g\to g-i\gamma$~\cite{LiuShiWang2024SciPost}.

In the present AGD setting, the \emph{short} scale $\ell$ arises because multiplying the spectrum by a finite window is equivalent, in the time domain, to convolving the \emph{absolute} autocorrelation with a sinc-like kernel of width $\sim \ell$.
Therefore, $\ell$ is best interpreted as a resolution-limited (high-frequency/short-time) correlation scale rather than an intrinsic decoherence time. A closely analogous role of an UV cutoff appears in nonequilibrium-condensate correlation theory: first-order correlations are often evaluated with an explicit UV cutoff to control short-distance/short-time behavior~\cite{DeLeeuwStoofDuine2014PRA}, and classical-field/SPGPE approaches implement the cutoff explicitly by splitting the field into a coherent region below an energy cutoff and an incoherent reservoir above it~\cite{KrauseBradley2026Evaporative}.

In contrast, the \emph{long} scale $\varrho$ is governed by the width and structure of the spectral core of $\hat S_{\rm AGD}$ and therefore reflects intrinsic low-frequency physics (slow collective modes and phase-noise accumulation), rather than the high-frequency cutoff. This parallels weakly nonequilibrium atomic condensates, where the coherence time is set by long-time phase dynamics rather than by UV regularization. A representative benchmark is the kinetic-theory result for condensate phase spreading, $\operatorname{Var}[\hat\varphi(t)]\simeq At^{2}+Bt+C$ at long times~\cite{SinatraCastinWitkowska2009PRA}.

From this viewpoint, Fig.~\ref{fig:agd_autocorr_two_scales} provides the following decomposition: (i) the bandwidth-induced graining sets the short time $\ell$; (ii) the AGD spectral core sets the long-time $\varrho$; and (iii) in the near-DSR regime, the energy-dependent scaling factor $\xi(\tilde E)$ (Eq.~(\ref{eq:scaling_corr})) primarily rescales the \emph{absolute} correlation amplitude, while the \emph{normalized} coherence remains nearly shape-invariant. This interpretation is fully consistent with the two-scale ``microstate'' picture: $\ell$ sets the effective temporal granularity of elementary degrees of freedom, while $\varrho$ sets the collective coherence time of their composite near the AGD--DSR line.

For driven optical condensates (polaritons/photons), finite lifetime and pump--reservoir coupling directly control damping and long-time coherence~\cite{CarusottoCiuti2013,DeLeeuwStoofDuine2014PRA}. For atomic condensates, engineered loss channels (including two-body loss) provide a tunable route to open-condensate behavior~\cite{LiuShiWang2024SciPost}, while the coherent/incoherent partition intrinsic to SPGPE/c-field methods makes the role of a cutoff scale explicit~\cite{KrauseBradley2026Evaporative}.
In this broader context, the near-DSR AGD regime offers a tractable example where the \emph{shape} of $g^{(1)}$ is governed by a nearly self-similar spectral core, while the \emph{strength} of first-order coherence is governed by an energy-dependent prefactor, separating ``coherence shape'' from ``coherence strength''.

\textbf{Self-similarity and energy scaling in correlation space.}
Along the near-resonant AGD--DSR paths (fixed $\tilde\chi$ and $C\simeq-1/\tilde\chi$), Fig.~\ref{fig:agd_autocorr_two_scales}(a) demonstrates that the windowed envelope spectra are approximately self-similar:
\[
\hat S^{(\mathrm{cap})}_{\mathrm{AGD}}(\tilde\Omega;\tilde E)\approx \xi(\tilde E)\,f(\tilde\Omega),\qquad
\int_{0}^{\tilde\Omega_{\mathrm{cap}}}\xi(\tilde E)\,f(\tilde\Omega)\,d\tilde\Omega=\tilde E,
\]
where $f(\tilde\Omega)$ is nearly $\tilde E$-independent (unit-peak shape) and the scalar factor $\xi(\tilde E)$ carries most of the energy variation.
Using Eq.~(\ref{eq:R1_AGD_def}) with $\hat S_{\mathrm{AGD}}\to \hat S^{(\mathrm{cap})}_{\mathrm{AGD}}$ then yields
\begin{equation}
R^{(1)}_{\mathrm{AGD,cap}}(\tilde\tau;\tilde E)\approx \xi(\tilde E)\,R^{(1)}_{f,\mathrm{cap}}(\tilde\tau),\qquad
g^{(1)}_{\mathrm{AGD,cap}}(\tilde\tau;\tilde E)\approx g^{(1)}_{f,\mathrm{cap}}(\tilde\tau),
\label{eq:scaling_corr}
\end{equation}
so that the normalized coherence function is nearly invariant with respect to $\tilde E$, whereas the absolute
correlation amplitude scales with $\xi(\tilde E)$. This is the time-domain analogue of the spectral self-similarity in Fig.~\ref{fig:agd_autocorr_two_scales}(a).

The two-scale structure predicted by the windowed-correlation representation (Eqs.~(\ref{eq:R1_convolution}--\ref{eq:rho_potential_AGD})) is made explicit in Fig.~\ref{fig:agd_autocorr_two_scales}.
The fine structure around $\tilde\tau=0$ is set by the finite bandwidth $\tilde\Omega_{\rm cap}$ through the sinc kernel, which defines the short \emph{graining} time $\ell\sim \pi/\tilde\Omega_{\rm cap}$, whereas the broad pedestal is controlled by the intrinsic width of the self-similar spectral core (the collective scale $\varrho$). Then, Eq.~(\ref{eq:scaling_corr}) explains the key qualitative separation: energy variation primarily rescales the absolute correlation amplitude
($R^{(1)}_{\mathrm{cap}}\propto \xi(\tilde E)$),
while the normalized coherence $g^{(1)}_{\mathrm{cap}}$ remains nearly shape-invariant,
providing a direct time-domain analogue of the near-DSR spectral self-similarity in Fig.~\ref{fig:fig5}(b).

Following \cite{kalashnikov2025energy}, one may interpret the separation $\ell\ll\varrho$ as the appearance of
``microstates'' (quasiparticles) of characteristic scale $\ell$ confined within a collective potential of scale $\varrho$.
A natural qualitative estimate for the effective number of microstates is then $N_{\rm eff}\sim \varrho/\ell$.
In AGD, the wing amplitude is controlled by the factor $(1+C\tilde\chi)^{-2}$ in the asymptotic tail (\ref{eq:AGDspa2}),
so approaching the AGD--DSR line enhances the contribution of high-frequency components and thereby increases
the scale separation, providing a route to the ``microstate multiplication'' scenario.
This suggests that a statistical-thermodynamic description \cite{picozzi2007towards,picozzi2009thermalization,picozzi2014optical,wu2019thermodynamic} of strongly chirped AGD DSs near DSR can be developed
in close analogy with the NGD case \cite{kalashnikov2025energy} \footnote{A fully closed-form expression for $R^{(1)}_{\rm AGD}(\tilde\tau)$ for the complete AGD envelope (\ref{eq:AGDspa1})
is generally unwieldy, but Eqs.~\eqref{eq:R1_AGD_def}--\eqref{eq:scaling_corr} provide a controlled analytic framework: the two-scale structure is governed by the effective bandwidth (or observation window) $\tilde\Omega_{\rm cap}$ and by the
core bandwidth $\tilde\Omega_{\rm core}$, while the near-DSR energy scalability manifests primarily through the scaling factor
$\xi(\tilde E)$.}.

\section{Conclusion}
We developed an adiabatic (strong-chirp) description of DSs of the cubic--quintic CGLE that treats NGD and AGD within a single conceptual framework. The resulting algebraic solvability and admissibility criteria naturally separate the two chirped branches and allow their
physically relevant existence domains to be defined in a compact master-diagram form, clarifying how the strongly chirped DSs persist under the dispersion-sign reversal and which constraints terminate the single-pulse states.

A central outcome is a transparent distinction between DSR mechanisms in the two dispersion regimes. In NGD, the DSR-type energy scalability appears as the natural scalable edge of the chirped branch under weak
dissipation, and the master diagram can be read as the boundary of the vacuum-stable domain. In AGD, by contrast, the DSR-type behavior is not generic. It becomes accessible only when the resonance locus is reachable inside the physically admissible strongly chirped AGD window, which in practice requires sufficiently strong quintic (saturable) SPM. That provides a concrete criterion for when the AGD energy scalability is expected and when it is prohibited by existence constraints, even before dynamical instabilities are considered.

Characterizing the DS spectra, the stationary-phase construction yields closed analytic predictions for both regimes. In AGD, the envelope lacks a strict cut-off and forms a structured core. Near the AGD-DSR locus, the unit-peak windowed spectral cores become weakly energy dependent, so that energy variation is captured predominantly by an overall scaling factor, while the normalized core shape remains nearly invariant. In general, the coherent stationary-phase field contains two dominant contributions whose interference provides a natural analytic route to the emergence of two- and three-horn spectra.

Interpreting the windowed spectrum as the relevant observable also leads to a simple coherence picture. The corresponding first-order autocorrelation decomposes into two correlation scales: a short-scale set by the dissipation-constrained spectral window and a long-scale controlled by the intrinsic spectral core. In the near-resonant regime, increasing energy primarily rescales the magnitude of the autocorrelation, whereas the normalized coherence retains
an approximately invariant shape when expressed in the properly scaled short- and long-time variables. This two-scale structure
provides an experimentally testable signature of coherence-scale decoupling in strongly chirped AGD DSs.

Finally, the separation between the short ``granularity'' scale and the long collective scale supports a thermodynamic/microstate
viewpoint: approaching the resonance enhances the scale separation. It can be interpreted as increasing the effective number of
accessible microstates, suggesting an entropy-driven tendency toward multi-pulse states with a DS energy growth. In this sense, strongly chirped DSs provide an analytically tractable bridge between ultrafast laser physics, nonlinear dynamical systems, and the broader language of weak-dissipative quantum condensates.

\section*{Acknowledgements}
  This work was supported by Norges Forskningsr{\aa}d (\#303347 (UNLOCK), \#326503 (MIR)), and ATLA Lasers AS.

\appendix
\section*{Appendix A. Saturable gain cavity maps and the interpretation of $\Sigma$ in master-diagram coordinates}

In the normalized description used throughout this work, the dimensionless net-loss parameter is $\Sigma \;=\; \frac{\zeta\,\sigma}{\kappa}$,
where $\sigma$ is the saturated net-loss (or saturated excess loss), and $\kappa,\zeta$ are, respectively, the cubic and quintic SAM coefficients in Eq.~(\ref{CGLE}). In the master-diagram construction \cite{kalashnikov2025energy}, $\Sigma$ is treated as an independent control parameter defining a stationary DS existence for given $(C,\tilde\chi,\Sigma)$.

In a cavity map with saturable gain, however, $\sigma$ is typically \emph{not} an independent parameter. Instead, $\sigma$ is updated self-consistently from the evolving pulse energy each round-trip, for example, by a stiffness law of the form
\begin{equation}
\sigma_{n+1}\;=\;(1-r)\sigma_n \;+\; r\,\delta\!\left(\frac{E_n}{E_{\rm cw}}-1\right),
\label{eq:sigmaClamp_app}
\end{equation}
or by a saturable-gain law $g(E)=g_0/(1+E/E_{\rm sat})$ combined with the fixed unsaturated losses.
In either case, the steady-state soliton selects a fixed operating point $(E,\sigma)$. Here, $n$ is the round-trip (map-iteration) index, $E_n$ is the pulse energy at round trip $n$
(e.g., $E_n=\int |a_n(t)|^2\,dt$ in the same units as $E_{\rm cw}$ or $E_{\rm sat}$). Here, $E_{\rm cw}$ and $E_{\rm sat}$ are the continuous-wave or gain-saturation energy, respectively; $\sigma_n$ is the (saturated) net-loss at round trip $n$
(with $\sigma>0$ meaning a vacuum-stable net loss and $\sigma<0$ meaning a vacuum-unstable net gain).
$r\in(0,1]$ is the relaxation/update factor controlling how fast $\sigma_n$ follows the instantaneous value
($r=1$ for instantaneous, smaller $r$ for slower);
$\delta$ is the stiffness of the energy-to-loss feedback in $\sigma_{n}(E)$, and $g_0$ is the small-signal (unsaturated) gain parameter in an explicit saturable-gain model (pump-controlled). 

Thus, $\Sigma$ is an output of the steady state rather than a scanned input. As a result, one may observe that $\sigma$ (and therefore $\Sigma$) depends only weakly on the stiffness $\delta$, on the pump proxy, or even moderately on $C$. These parameters primarily change how the map converges, while the stationary pulse balance selects the net-loss level required for equilibrium.

Eq.~(\ref{norm}) shows that $\Sigma$ depends on the ratio $\zeta/\kappa$ as well as on $\sigma$.
In the gain-saturable simulations, the same coefficients $\kappa,\zeta$ also enter the energy normalization (e.g., through the definition of the saturation energy $E_{\rm sat}$ or its counterpart like $E_{\rm cw}$).
As a consequence, the varying $\zeta$, while keeping the cavity-map structure unchanged, changes the mapping between ``pump'' parameters and the dimensionless operating point.
The simulation then readjusts the steady-state energy so that the saturation rule \eqref{eq:sigmaClamp_app} produces a new $\sigma$.
It is therefore common to observe partial compensation: $\sigma$ shifts so that the combination $\zeta\sigma/\kappa$ (i.e., $\Sigma$) remains approximately fixed. Thus, the cavity map selects (approximately) one effective isogain slice in a master diagram.

In the master-diagram studies of chirped DSs, the stability/existence regions are often visualized as being ``filled'' by isogain curves \cite{kalashnikov2025energy}: each curve corresponds to a different saturated net-loss level (equivalently, a different pump/unsaturated-loss operating point), and therefore to a different value of $\Sigma$. In a single fixed cavity-map run, only one (or a very narrow set of) operating points is realized, so the numerically observed solutions naturally cluster near an effectively constant $\Sigma$.
To reproduce the filled master-diagram picture numerically, one must introduce an \emph{independent} parameter that shifts the gain--loss balance without being slaved to only the pulse energy. For instance, one may use:

\begin{equation}
\sigma_{n+1}=(1-r)\sigma_n+r\left[\sigma_0+\delta\!\left(\frac{E_n}{E_{\rm cw}}-1\right)\right],
\end{equation}
and scan a $\sigma_0$ (an offset or baseline net-loss) at fixed $(\kappa,\zeta,\alpha,\beta,\gamma,\chi)$.
This produces a family of steady states with different $\sigma$ and hence different $\Sigma$. Or one may use an explicit saturable-gain model,
\begin{equation}
\sigma(E)=L-\frac{g_0}{1+E/E_{\rm sat}},
\end{equation}
and scan the small-signal gain $g_0$ (pump) or/and the unsaturated loss $L$.
This allows generating a family of operating points with different $\Sigma$ and populating the master-diagram region. Without such an additional independent knob, a gain-clamped map is expected to yield an approximately single-$\Sigma$ operating line even if $E$ varies strongly.
\label{app:gainclamp_sigmaSigma}

\appendix
\section*{Appendix B. Uniform Airy approximation at the NGD spectral edge}

Let us write the (dimensionless) spectral amplitude as a rapidly oscillatory Fourier–type integral
\begin{equation}
\widehat{a}(\tilde\Omega)=\int_{-\infty}^{\infty} A(t)\,e^{\tfrac{i}{\varepsilon}\Psi(t;\tilde\Omega)}\,dt,\qquad 
\Psi(t;\tilde\Omega)=\Phi(t)-\tilde\Omega\,t,\quad \Phi'(t)=\tilde\Omega_{\rm inst}(t).
\end{equation}
\noindent Here $\varepsilon\ll 1$ is the adiabatic (large-chirp) parameter. With the normalization of Sec.~2 (cf.\ Eq.~(\ref{norm})), it is inversely proportional to the chirp magnitude, so SPA/CFU expansions are controlled by $\varepsilon\to 0$.

Inside the band, the SPA condition $\partial_t\Psi=0$ has two real roots $t_\pm(\tilde\Omega)$, which coalesce at the cut-off $|\tilde\Omega|=\tilde\Delta$. Thus, the standard two-saddle SPA loses uniformity. Near the turning point $(t_*,\tilde\Omega=\tilde\Delta)$, a \emph{Chester--Friedman--Ursell (CFU)} change of variables reduces the phase to the Airy canonical cubic (see, e.g., Wong~\cite{Wong2001}):

\begin{equation}
\Psi(t;\tilde\Omega)=\Psi_*+\frac{\varsigma}{3}s^3-\varsigma^{1/3}\upsilon\,s+O(\varepsilon^{2/3}),\qquad 
\varsigma\equiv\tfrac12\big|\tilde\Omega''_{\rm inst}(t_*)\big|,\quad 
\upsilon=\varepsilon^{-2/3}\vartheta\,(\tilde\Delta-\tilde\Omega),\ \vartheta=\varsigma^{-2/3}.
\end{equation}

Freezing the factor $\mathcal{G}(s;\varepsilon)$ at the turning point gives the leading uniform approximation
\begin{equation}
\widehat{a}_{\rm unif}(\tilde\Omega)=e^{i(\Psi_*/\varepsilon+\pi/6)}\left(\frac{2\varepsilon}{\varsigma}\right)^{1/3}
\!\left\{a_0\,\mathrm{Ai}(\upsilon)+\varepsilon^{1/3}a_1\,\mathrm{Ai}'(\upsilon)\right\}+O(\varepsilon^{4/3}),
\end{equation}
with $a_0=\mathcal{G}_*$ and a computable $a_1$ (a ``transport'' correction). Keeping only the $\mathrm{Ai}$ term already yields a uniform error $O(\varepsilon^{2/3})$ but restores finiteness at the edge.

As $\upsilon\!\to\!0$ (the cut-off), $\mathrm{Ai}(0)=3^{-2/3}/\Gamma(2/3)$ gives a finite nonzero edge amplitude. For $\upsilon\!\to\!+\infty$ (outside the band), one obtains the exponentially small tail $\,\mathrm{Ai}(\upsilon)\sim \tfrac{1}{2\sqrt{\pi}}\upsilon^{-1/4}e^{-\tfrac23\upsilon^{3/2}}$.
Inside the band ($\upsilon\!\to\!-\infty$), the uniform expression reproduces the interference of the two interior stationary points.

With our normalization $\widehat{S}(\tilde\Omega)=\mathcal{P}\,|\widehat{a}(\tilde\Omega)|^2$ ($\mathcal{P}$ as in the main text), a robust one-line uniformization that exactly matches the analytic edge value from the SPA formula is
\begin{equation}
\widehat{S}_{\rm unif}(\tilde\Omega)\;=\;\widehat{S}_{\rm edge}\,
\bigg[\frac{\mathrm{Ai}\!\big(\upsilon(\tilde\Omega)\big)}{\mathrm{Ai}(0)}\bigg]^2,\qquad
\upsilon(\tilde\Omega)=\varepsilon^{-2/3}\vartheta\,(\tilde\Delta-\tilde\Omega)\;,
\label{eq:airy-min}
\end{equation}
with the symmetric form obtained by replacing $(\tilde\Delta-\tilde\Omega)$ by $(\tilde\Delta-|\tilde\Omega|)$.
Here $\widehat{S}_{\rm edge}$ is the finite cut-off value from the closed NGD expression. This replacement removes the $0/0$ singularity at $|\tilde\Omega|=\tilde\Delta$, smooths the cut-off, and agrees with the interior SPA away from the edge.
\appendix

\appendix
\section*{Appendix C. Autocorrelation function for a strongly chirped AGD DS}
\label{app:acf-agd}

\subsection*{C.1. Definitions and windowed spectrum}
In the AGD regime ($\beta<0$, $C<0$, $\tilde\chi>0$, $\tilde\Delta_-^2<0$), the SPA yields
the envelope spectrum $\hat S_{\rm AGD}(\tilde\Omega)$ given by Eq.~(\ref{eq:AGDspa1}). This spectrum has no strict cut-off and decays
as $\hat S_{\rm AGD}(\tilde\Omega)\sim|\tilde\Omega|^{-3}$, Eq.~(\ref{eq:AGDspa2}), so the energy is finite and can be written as
Eq.~(\ref{eq:enAGD}).

In practice (finite spectral dissipation, measurement bandwidth, or the finite plotting window used in Fig.~\ref{fig:fig5}(b)),
one observes a windowed spectrum
\begin{equation}
\hat S_{\rm AGD}^{({\rm cap})}(\tilde\Omega)
=
\hat S_{\rm AGD}(\tilde\Omega)\,H(\tilde\Omega_{\rm cap}-|\tilde\Omega|),
\label{eq:S_cap}
\end{equation}
where $H$ is the Heaviside function and $\tilde\Omega_{\rm cap}$ is the effective observation bandwidth.

The (first-order) field autocorrelation is defined by the cosine transform
\begin{equation}
R^{(1)}_{\rm cap}(\tilde\tau)
=
\frac{1}{\pi}\int_{0}^{\infty}\hat S_{\rm AGD}^{({\rm cap})}(\tilde\Omega)\cos(\tilde\Omega\tilde\tau)\,d\tilde\Omega,
\qquad
g^{(1)}_{\rm cap}(\tilde\tau)=\frac{R^{(1)}_{\rm cap}(\tilde\tau)}{R^{(1)}_{\rm cap}(0)}.
\label{eq:R1_def}
\end{equation}
(If the field fluctuations are close to Gaussian, the intensity autocorrelation follows from a Siegert-type relation:
$g^{(2)}(\tilde\tau)=1+|g^{(1)}(\tilde\tau)|^2$.)

\subsection*{C.2. Convolution form and the short correlation scale}
Let $\mathcal{F}^{-1}$ denote the inverse Fourier transform in $\tilde\Omega$.
Since multiplication in frequency corresponds to convolution in time, Eq.~\eqref{eq:S_cap} implies
\begin{equation}
R^{(1)}_{\rm cap}(\tilde\tau)
=
R^{(1)}_{\infty}(\tilde\tau)\ast K_{\rm cap}(\tilde\tau),
\qquad
K_{\rm cap}(\tilde\tau)=\frac{\tilde\Omega_{\rm cap}}{\pi}\,\mathrm{sinc}\!\left(\tilde\Omega_{\rm cap}\tilde\tau\right),
\label{eq:conv}
\end{equation}
where $R^{(1)}_{\infty}$ is the correlation corresponding to the unwindowed spectrum $\hat S_{\rm AGD}$ and
$\mathrm{sinc}(x)\equiv\sin(x)/x$.
Thus, the spectral windowing introduces a \emph{short} correlation scale
\begin{equation}
\ell \sim \frac{\pi}{\tilde\Omega_{\rm cap}},
\label{eq:ell_def}
\end{equation}
which is the characteristic width of $K_{\rm cap}$.
This is the direct AGD analogue of the ``$\mathrm{sinc}$-graining'' mechanism and the microscopic time scale
identified for strongly chirped DSs in Appendix~B of Ref.~\cite{kalashnikov2025energy}.

\subsection*{C.3. Long correlation scale from the spectral core}
The \emph{long} correlation scale is controlled by the spectral core of $\hat S_{\rm AGD}$.
A convenient operational definition is to introduce a core bandwidth $\tilde\Omega_{\rm core}$ by an energy fraction
\begin{equation}
\int_{0}^{\tilde\Omega_{\rm core}}\hat S_{\rm AGD}(\tilde\Omega)\,d\tilde\Omega
=
(1-\varepsilon)\int_{0}^{\infty}\hat S_{\rm AGD}(\tilde\Omega)\,d\tilde\Omega,
\qquad 0<\varepsilon\ll 1,
\label{eq:Ocore_def}
\end{equation}
and set
\begin{equation}
\varrho \sim \tilde\Omega_{\rm core}^{-1}.
\label{eq:rho_def}
\end{equation}
Then $|g^{(1)}_{\rm cap}(\tilde\tau)|$ typically exhibits a broad pedestal of width $\sim\varrho$,
while the $\mathrm{sinc}$-kernel in Eq.~\eqref{eq:conv} produces a much narrower structure of width $\sim\ell$,
so that a two-scale separation $\ell\ll\varrho$ emerges.

For the analytic estimations, one may approximate the core of the AGD spectrum by a Lorentzian
$\hat S_{\rm core}(\tilde\Omega)\approx A\,(1+\tilde\Omega^2/\Gamma^2)^{-1}$,
which gives $R_{\rm core}(\tilde\tau)\propto e^{-\Gamma|\tilde\tau|}$ and hence $\varrho\sim 1/\Gamma$.
Substitution into Eq.~\eqref{eq:conv} yields the explicit two-scale convolution
\begin{equation}
R^{(1)}_{\rm cap}(\tilde\tau)
\approx
\int_{-\infty}^{\infty} e^{-|t|/\varrho}\,
\frac{\tilde\Omega_{\rm cap}}{\pi}\,\mathrm{sinc}\!\left(\tilde\Omega_{\rm cap}(\tilde\tau-t)\right)\,dt,
\label{eq:two_scale_model}
\end{equation}
which is structurally identical to the NGD strong-chirp autocorrelation formula of Ref.~\cite{kalashnikov2025energy}.

\subsection*{C.4. Contribution of the algebraic AGD tail}
The AGD envelope has an algebraic wing $\hat S_{\rm AGD}(\tilde\Omega)\sim K/|\tilde\Omega|^{3}$ with
\begin{equation}
K \propto \frac{1}{(1+C\tilde\chi)^2\sqrt{\tilde\chi}},
\label{eq:K_tail}
\end{equation}
as follows from Eq.~(\ref{eq:AGDspa2}). Let $\tilde\Omega_0$ denote a matching frequency beyond which the $|\tilde\Omega|^{-3}$ asymptotic is accurate. The tail contribution to $R^{(1)}_{\rm cap}$ is then
\begin{equation}
R_{\rm wing}(\tilde\tau)
\approx
\frac{K}{\pi}\int_{\tilde\Omega_0}^{\tilde\Omega_{\rm cap}}
\frac{\cos(\tilde\Omega\tilde\tau)}{\tilde\Omega^{3}}\,d\tilde\Omega.
\label{eq:R_wing_def}
\end{equation}
Expanding $\cos(\tilde\Omega\tilde\tau)$ for $|\tilde\tau|\ll \tilde\Omega_{\rm cap}^{-1}$ gives
\begin{equation}
R_{\rm wing}(\tilde\tau)
=
R_{\rm wing}(0)
-\frac{K}{2\pi}\,\tilde\tau^{2}\,\ln\!\Bigl(\frac{\tilde\Omega_{\rm cap}}{\tilde\Omega_0}\Bigr)
+\mathcal{O}(\tilde\tau^{4}),
\label{eq:R_wing_smalltau}
\end{equation}
showing that the curvature of the correlation peak depends logarithmically on the spectral cut-off.
Near the AGD--DSR line $1+C\tilde\chi\simeq 0$, the prefactor $K$ increases, enhancing the relative weight of high-frequency
components and, thus, amplifying the short-scale structure (i.e., increasing the separation between $\ell$ and $\varrho$).

\subsection*{C.5. Connection to ``microstates''}
Following Ref.~\cite{kalashnikov2025energy}, the emergence of two disparate scales $\ell\ll\varrho$ motivates the
interpretation of the strongly chirped DS as a composite of microscopic degrees of freedom (``microstates'') of characteristic
scale $\ell$ confined within a collective envelope of scale $\varrho$.
A crude estimate for the effective number of microstates is $N_{\rm eff}\sim \varrho/\ell$, which increases as one approaches
the AGD--DSR resonance (through the growth of the wing prefactor~\eqref{eq:K_tail} and/or through an increase in the effective
bandwidth $\tilde\Omega_{\rm cap}$ required to capture the spectrum). As was conjectured in \cite{kalashnikov2025energy}, the microstate multiplication (i.e, $N_{\rm eff}$ growth) increases the DS entropy and switches the DS temperature from a positive to a negative value, meaning a transit to a strongly nonequilibrium state and loss of DS stability due to multiple soliton generation.

\printcredits

\bibliographystyle{model1-num-names}

\bibliography{cas-refs}



\end{document}